\documentclass{aa}
\usepackage[fleqn]{amsmath}
\usepackage{graphicx}
\usepackage{dcolumn}
\usepackage{bm}
\usepackage{rotating}
\usepackage{threeparttable}
\usepackage{amsmath}
\usepackage{verbatim}

\usepackage{longtable}
\usepackage{mdwlist}
\usepackage{lscape}
\usepackage{natbib}
\usepackage{multirow}
\usepackage{upgreek}
\usepackage[colorlinks=true, pdfstartview=FitV, linkcolor=blue,citecolor=blue, urlcolor=blue]{hyperref}
\usepackage{wasysym}
\usepackage{txfonts}
\usepackage{gensymb}
\usepackage{morefloats}

\begin{document}

\bibliographystyle{plain}

\title{Estimating ages and metallicities of M31 star clusters from LAMOST DR6}
\titlerunning{M31 clusters from LAMOST DR6}
\authorrunning{Wang et al.}

\author{Shoucheng Wang\inst{1,2,3}  \and
          Bingqiu Chen\inst{2} \and
          Jun Ma\inst{1,3}}

\institute{Key Laboratory of Optical Astronomy, National Astronomical Observatories, Chinese Academy of Sciences, Beijing 100012, China e-mail: majun@nao.cas.cn;
\and
South-Western Institute for Astronomy Research, Yunnan University, Kunming, Yunnan 650091, P. R. China e-mail: bchen@ynu.edu.cn;
\and
School of Astronomy and Space Sciences, University of Chinese Academy of Sciences, Beijing 100049, China}

\abstract
{Determining the metallicities and ages of M31 clusters is fundamental to the study of the formation and evolution of M31 itself. The Large Sky Area Multi-Object Fiber Spectroscopic Telescope (LAMOST) has carried out a systematic spectroscopic campaign of clusters and candidates in M31.}
{We constructed a catalogue of 346 M31 clusters observed by LAMOST. By combining the information of the LAMOST spectra and the multi-band photometry, we developed a new algorithm to estimate the metallicities and ages of these clusters. }
{We distinguish young clusters from  old using random forest classifiers based on a empirical training data set selected from the literature. Ages
of young clusters are derived from the spectral energy distribution (SED) fits of their multi-band photometric measurements. Their metallicities are estimated by fitting their observed spectral principal
components extracted from the LAMOST spectra with those from the young metal-rich single stellar population (SSP) models. For old clusters we  built non-parameter  random forest models between the
spectral principal components and/or multi-band colours and the parameters of the clusters based on a training data set constructed from the SSP models. The ages and metallicities of the old clusters are then estimated
by fitting their observed spectral principal components extracted from the LAMOST spectra and multi-band colours from the photometric measurements with the resultant random forest models.
}
{We  derived parameters of 53 young and 293 old clusters in our catalogue. Our resultant parameters are in good agreement with those from the literature. The ages of $\sim$ 30 catalogued clusters and metallicities of  $\sim$ 40 sources are derived for the first time.
}
{}


\maketitle

\section{Introduction}

Star clusters in the Andromeda galaxy M31 are excellent tracers in  studies of the formation and evolution of M31 itself. In the past decades many works have been performed to identify and classify the clusters in M31. \citet{2000AJ....119..727B} presented a catalogue of 435 clusters and candidates in M31 from the literature. \citet{2004A&A...416..917G} identified 693 known and candidate globular clusters (GCs) in M31 in the 2MASS database and presented the Revised Bologna catalogue (RBC$\footnote{http://www.bo.astro.it/M31/}$) of M31 globular clusters and candidates. \citet{2012ApJ...752...95J} identified 601 clusters based on the data from Panchromatic Hubble Andromeda Treasury (PHAT; \citealt{2012ApJS..200...18D}). \citet{2013AJ....145...50D} searched M31 outer halo GCs using the image data from the Sloan Digital Sky Survey (SDSS; \citealt{2002AJ....123..485S}) and identified 18 new clusters. Based on the data from the Pan-Andromeda Archaeological Survey (PAndAS; \citealt{2009AAS...21330705M}), \citet{2014MNRAS.442.2165H} identified 59 GCs and 2 candidates in the halo of M31.

Determining the parameters, such as the metallicities and ages, of M31 clusters is fundamental to the study of the formation and evolution of their parent galaxy. Due to the large distance of M31 it is hard to resolve the member stars of the cluster. Thus, most studies interpret the cluster in M31 as a simple stellar population (SSP) and estimate its parameters by fitting its integrated colours or integrated spectrum with the SSP models (\citealt{1988ARA&A..26..199R,2003MNRAS.344.1000B,2009MNRAS.396..462K,2010MNRAS.404.1639V}). For example, \citet{2009AJ....137...94C,2011AJ....141...61C} presented a catalogue of over 100 young and 300 old clusters with parameters determined from high signal-to-noise (S/N) spectroscopic data. \citet{2012AJ....143...29M} determined the age and mass of B514 by comparing its spectral energy distribution (SED) with theoretical stellar population synthesis models. \citet{2013A&A...549A..60C} obtained ages and metallicities of 38 M31 and 41 Galactic GCs by comparing their observed integrated spectra with the SSP model spectra on a pixel-by-pixel basis.  \citet{2016AJ....152..208F} derived ages and metallicities of 22 confirmed M31 GCs and concluded that using a combination of spectroscopic and photometric data resulted more reliable parameters than those from either spectroscopic data only or photometric data only. \citet{2019A&A...623A..65W} determined the metallicities, ages, and masses of 53 GCs in the M31 outer halo by comparing the multi-colour photometry with stellar population synthesis models. \citet{2020IAUS..351..131N} derived star cluster ages and metallicities in M31 by employing stochastic star cluster models. Machine learning algorithms are also adopted to derive star cluster parameters, such as convolutional neural networks (CNNs), which are used in \citet{2019A&A...621A.103B} to develop a CNN-based algorithm capable of simultaneously deriving ages, masses, and sizes of star clusters directly from multi-band images.

The Large Sky Area Multi-Object Fiber Spectroscopic Telescope (LAMOST; \citealt{2012RAA....12.1197C}) is a quasi-meridian reflecting Schmidt telescope with 4000 fibers distributed in a field of view of $5^{\circ}$ in diameter. It is an ideal facility to carry out a systematic spectroscopic survey of M31 clusters and candidates. As a part of the LAMOST Spectroscopic Survey of the Galactic Anti-centre (LSS-GAC; \citealt{2014IAUS..298..310L}), the LAMOST M31/M33 survey targets observable objects in M31. Based  on LAMOST M31/M33 survey data that collected in the 2011, 2012, and 2013 observational seasons, \citet{2015RAA....15.1392C}  presented a catalogue of 908 objects observed by LAMOST in the vicinity fields of M31 and M33, of which 356 are likely GCs of various degrees of confidence. Based on the catalogue, \citet{2016AJ....152...45C}   estimated metallicities, ages and masses of 306 confirmed massive clusters by different methods, including the full spectrum fitting technique and the SED fits.

In this work we  present a catalogue of M31 clusters from LAMOST Data Release 6 (DR6$\footnote{http://dr6.lamost.org/}$; \citealt{2015RAA....15.1095L}). Compared to the catalogue of \citet{2016AJ....152...45C}, which used data from LSS-GAC DR2 \citep{2017MNRAS.467.1890X}, the new catalogue presented in the current work contains more objects. The LAMOST spectra from DR6 are processed with an improved pipeline. Moreover, we have also developed a new algorithm that combines the information of spectroscopic and photometric observations of the clusters to derive their metallicities and ages.

This paper proceeds as follows. In Section 2 we introduce our data set and the cluster sample selection. Then we describe the parameters determination methods in Section 3. We present our results in Section 4, and summarize in Section 5.

\section{Data}

\subsection{LAMOST DR6}
\begin{figure*}
\center{
\includegraphics[scale=0.45]{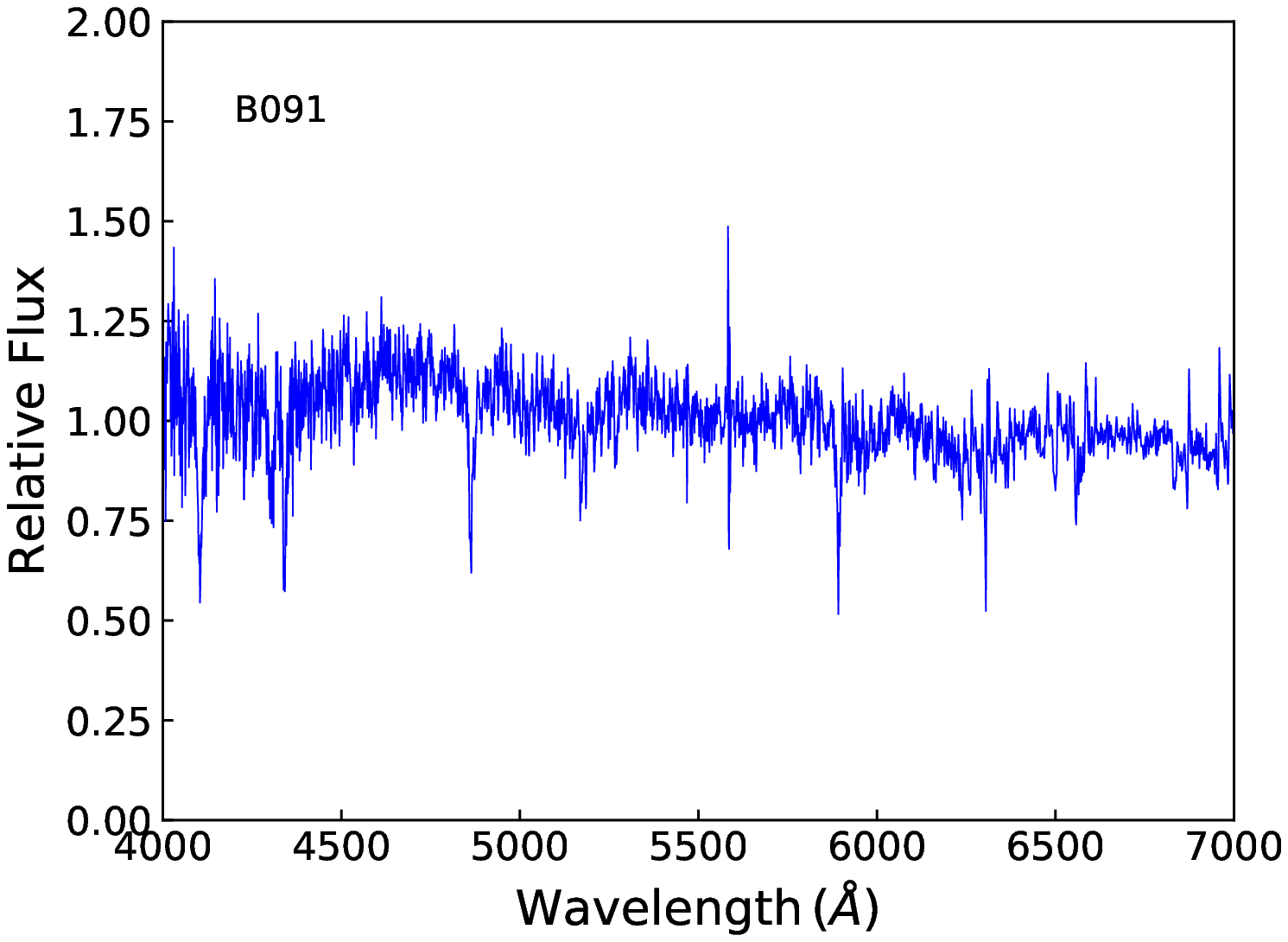}
\includegraphics[scale=0.45]{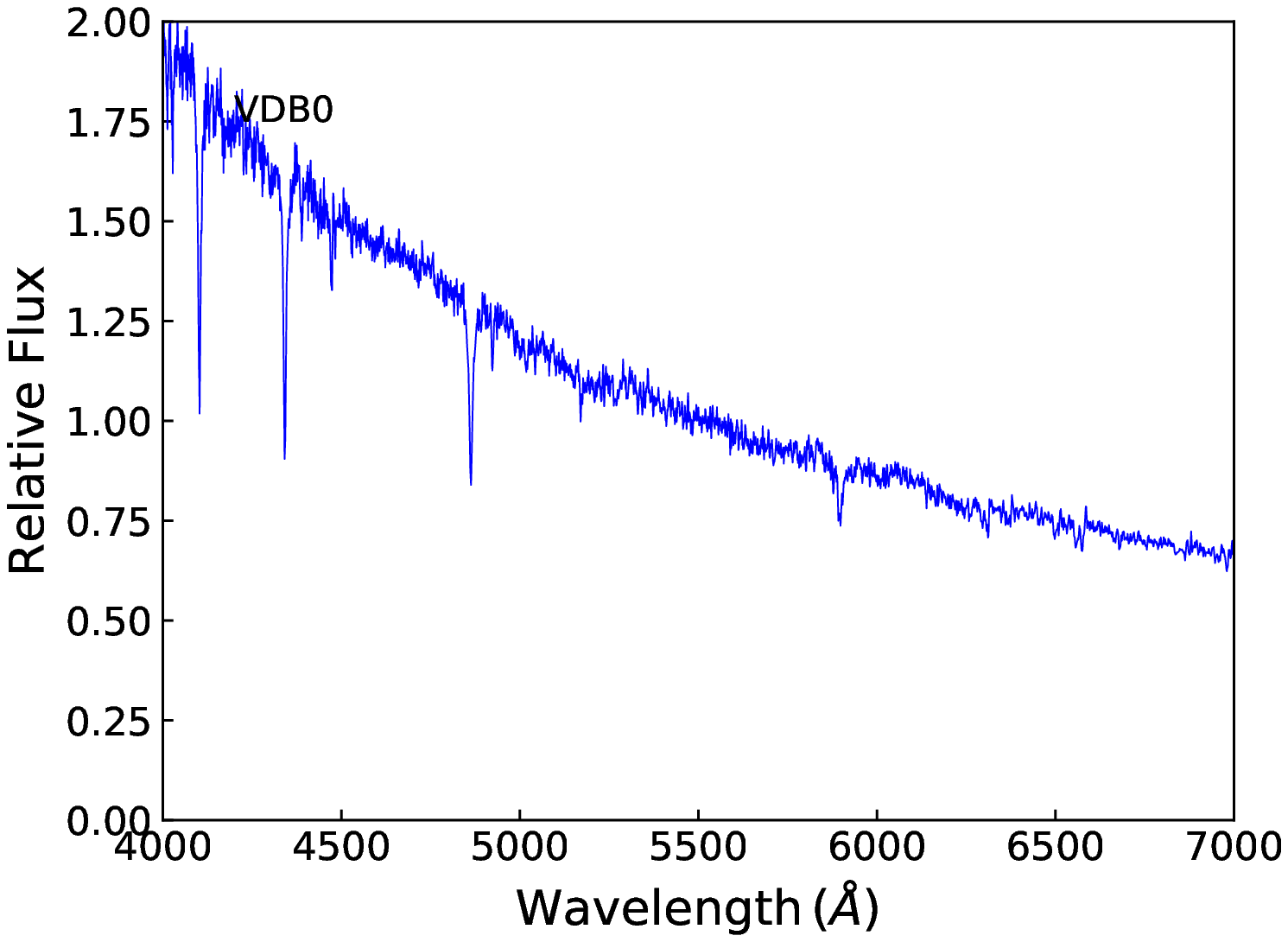}
\includegraphics[scale=0.45]{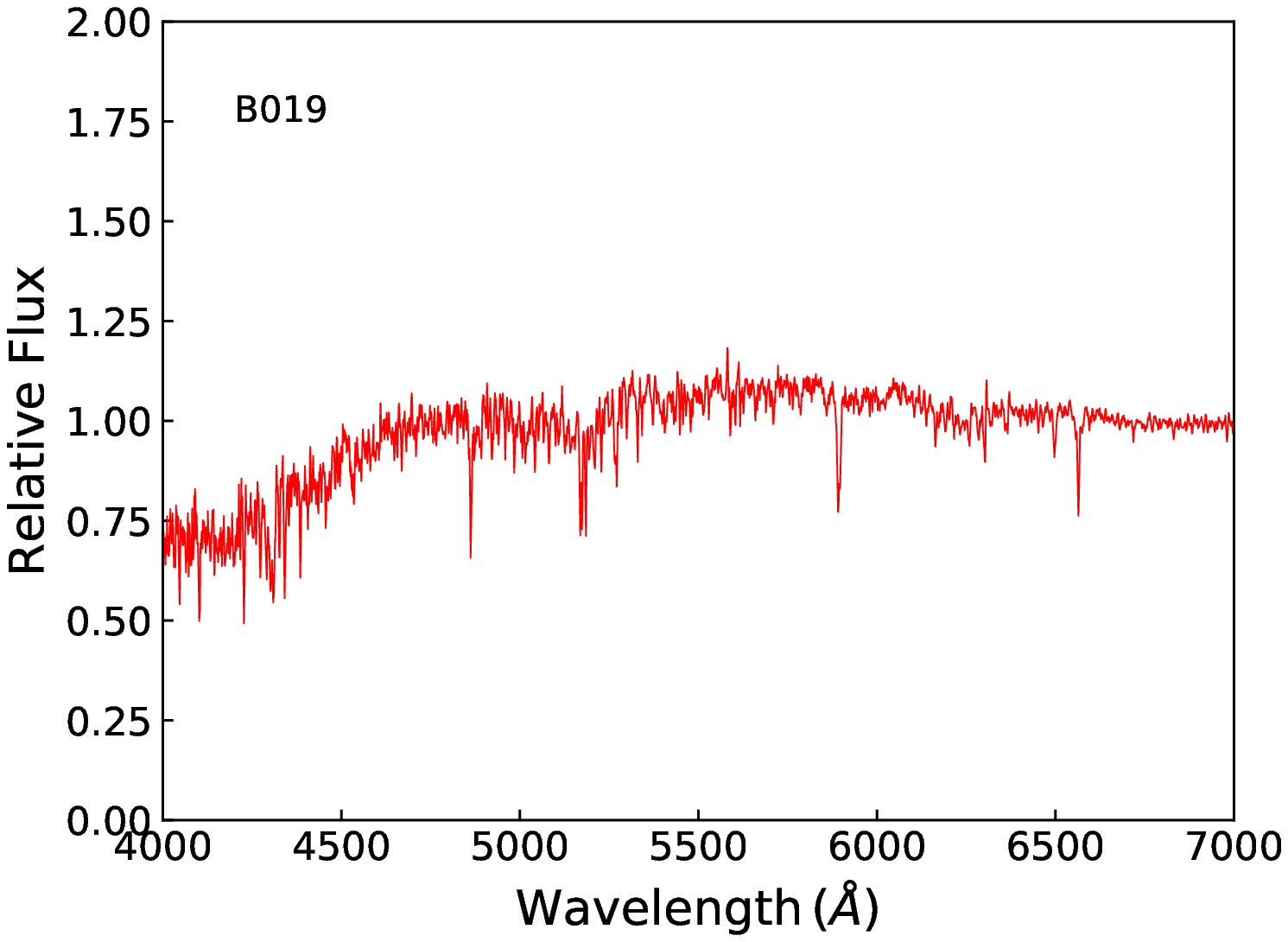}
\includegraphics[scale=0.45]{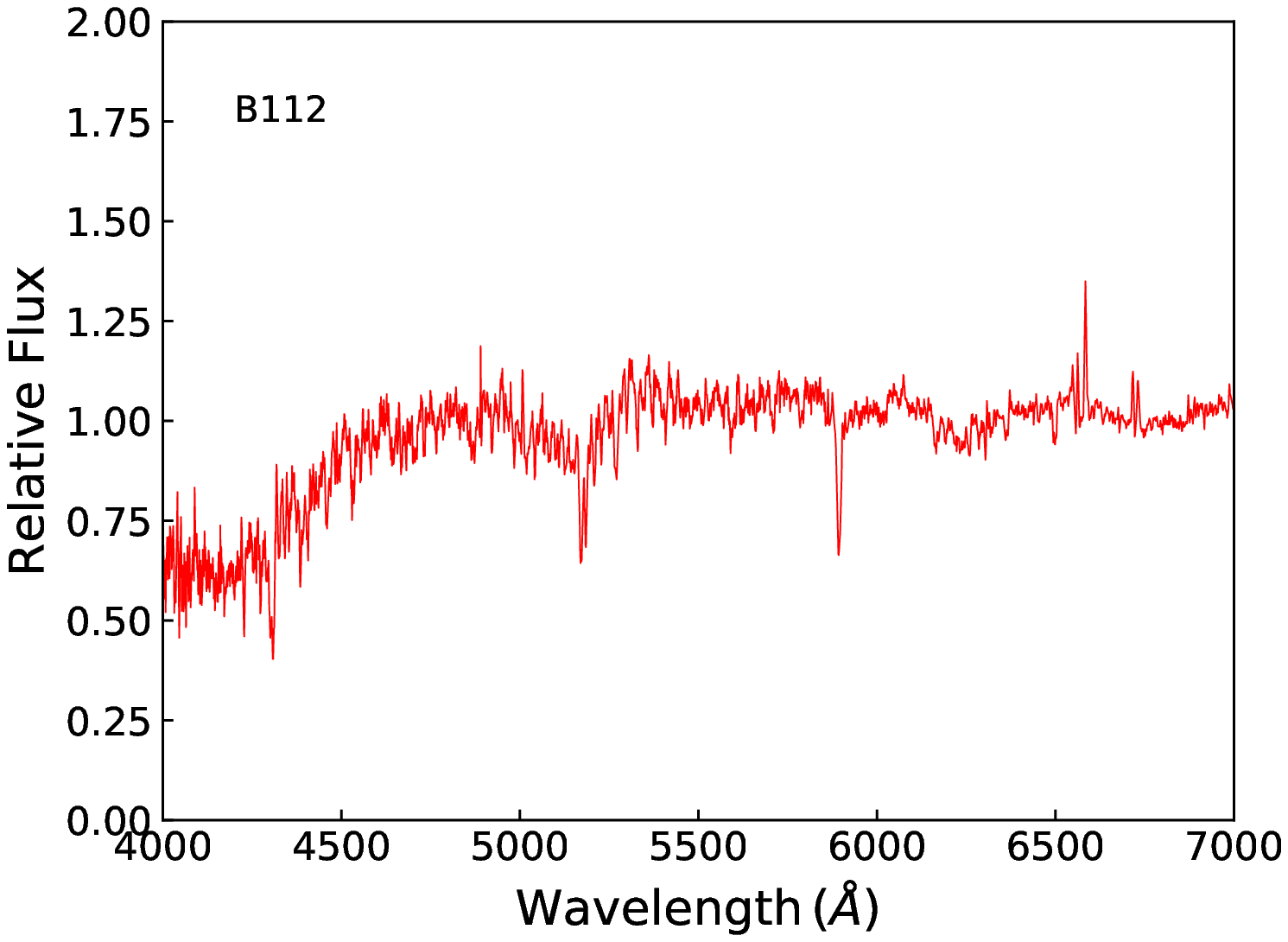}
}
\caption{
Example LAMOST spectra of four star clusters in our sample. The cluster names are indicated for each spectrum. The upper and lower panels show spectra of two young and two old clusters (in blue and red, respectively). The fluxes are normalized by median values of the individual clusters. To be consistent with our fitting wave band, only spectra in the wavelength range $4000$--$7000\,\rm\AA$ are plotted.
}
\label{fig:lamost}
\end{figure*}

Our work is based on the LAMOST DR6, which consists of two components, the low-resolution part and the medium-resolution part. The LAMOST M31/M33 is a low-resolution survey. In the current work we use only the low-resolution data which contains 9,919,106 calibrated spectra. The spectra cover the wavelength range of $3690\sim9100\,\rm\AA$ with a resolution of $R\sim1800$ at  $5500\,\rm\AA$. In the current work we use the 1D spectra of objects from LAMOST DR6, which are extracted by the LAMOST 2D and 1D pipeline \citep{2013IAUS..295..189Z}. Example spectra of four star clusters (two young and two old; see Section 4) are presented in Fig.~\ref{fig:lamost}.

\subsection{Sample selection}
To make  our catalogue of M31 clusters as complete and uncontaminated as possible, we collected the confirmed clusters from several previous works.
\begin{itemize}
        \item We started with the RBC (\citealt{2004A&A...416..917G}), which is a compilation of previously published catalogues. It includes 336 confirmed M31 GCs (class = 1).
        \item The catalogue from \cite{2012ApJ...752...95J} contains 601 confirmed clusters (denoted as `c') selected from the high spatial resolution images of the Panchromatic Hubble Andromeda Treasury (PHAT) covering one-third of the M31 disk at ultraviolet through near-infrared wavelengths.
        \item The catalogue from \citet{2013AJ....145...50D} contains 18 confirmed clusters in the halo of M31 selected from the SDSS data.
    \item The catalogue from \citet{2014MNRAS.442.2165H} contains 59 GCs in the halo of M31 selected from the Pan-Andromeda Archaeological Survey (PAndAS) survey.
        \item The catalogue from \cite{2015RAA....15.1392C} contains 306 star clusters observed by LAMOST M31/M33 survey since June 2014.
\end{itemize}

As a result, we   obtained a catalogue of 1233 confirmed M31 clusters. These objects are matched with the LAMOST DR6 with a radius of $3^{\prime\prime}$. Sources with LAMOST spectra having S/N in $g$ band lower than 5 or S/N in $i$ band lower than 10 are excluded. This leaves us 334 confirmed clusters.

In addition to the confirmed clusters, we  also collected 664 cluster candidates from the literature  listed above. These objects are also matched with LAMOST DR6. Again we   excluded sources with LAMOST spectra having S/N in $g$ band lower than 5 or S/N in $i$ band lower than 10. As in  the work of \citet{2006A&A...456..985G} and \citet{2015RAA....15.1392C}, we classify the candidates having radial velocities $V_{r}$ lower than $-150\,\rm km\,s^{-1}$ as bona fide clusters in the current work. As a result, we are left with 12 clusters with $V_{r}<-150\,\rm km\,s^{-1}$.

\begin{figure}
\center{
\includegraphics[scale=0.37]{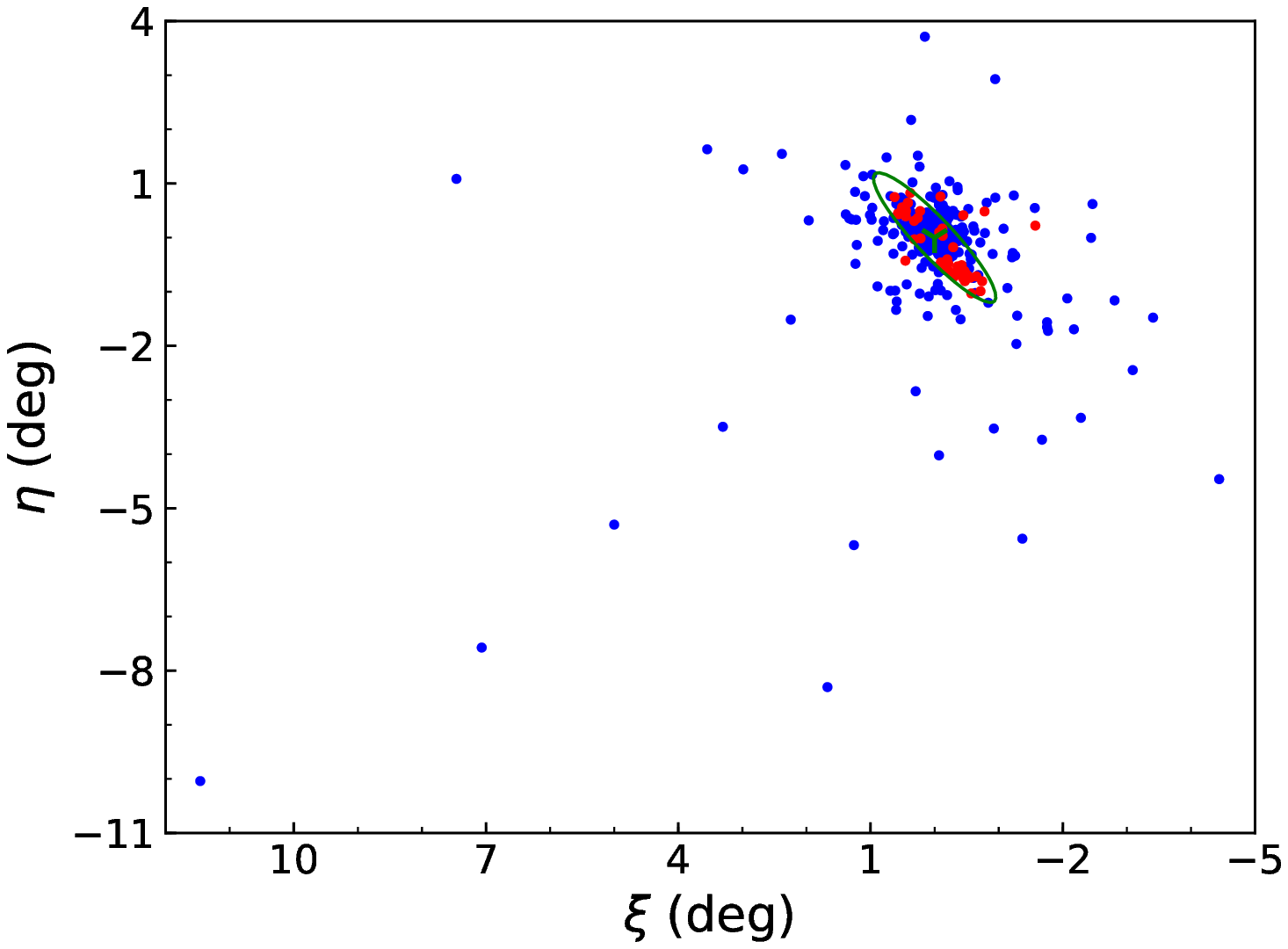}
\includegraphics[scale=0.37]{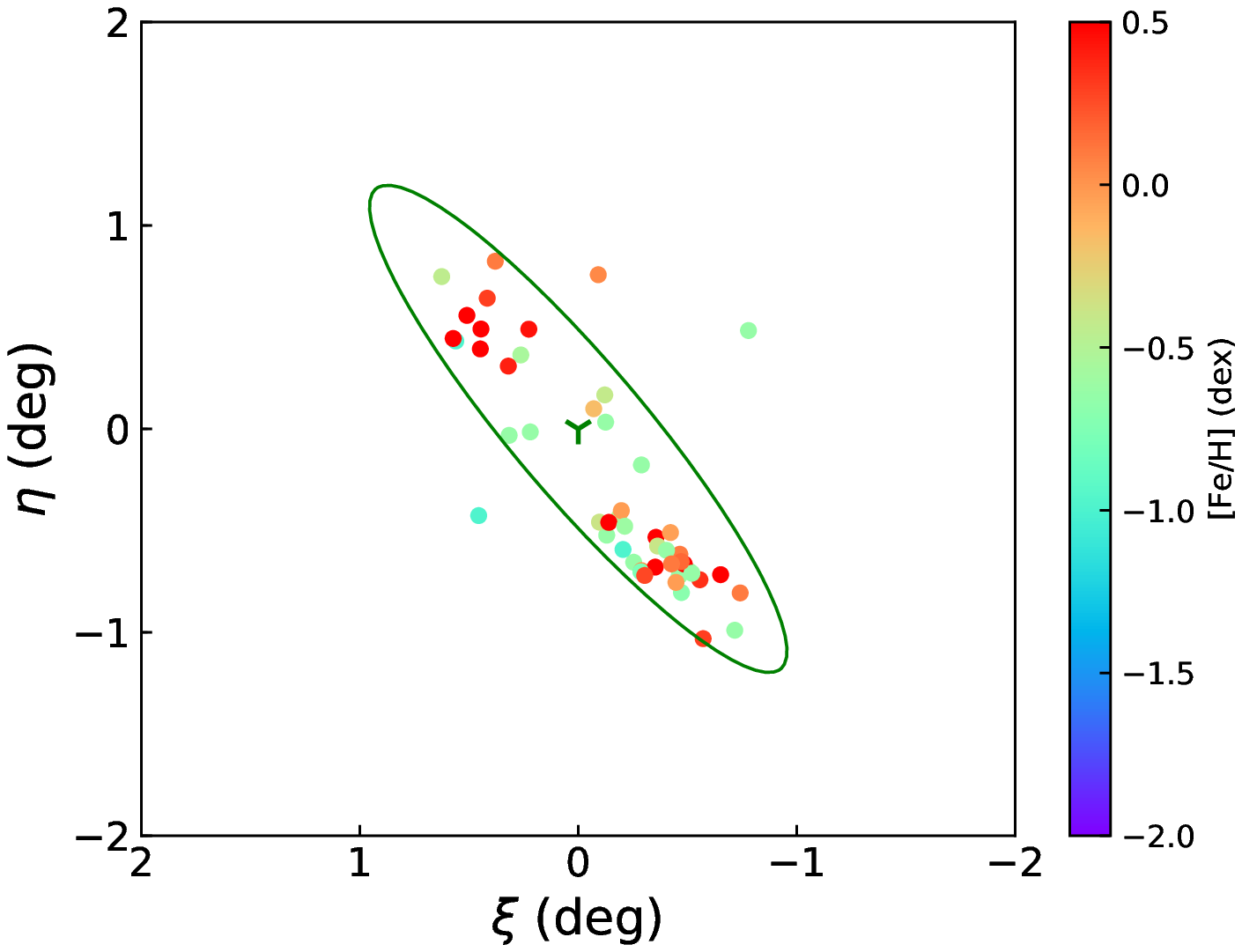}
\includegraphics[scale=0.37]{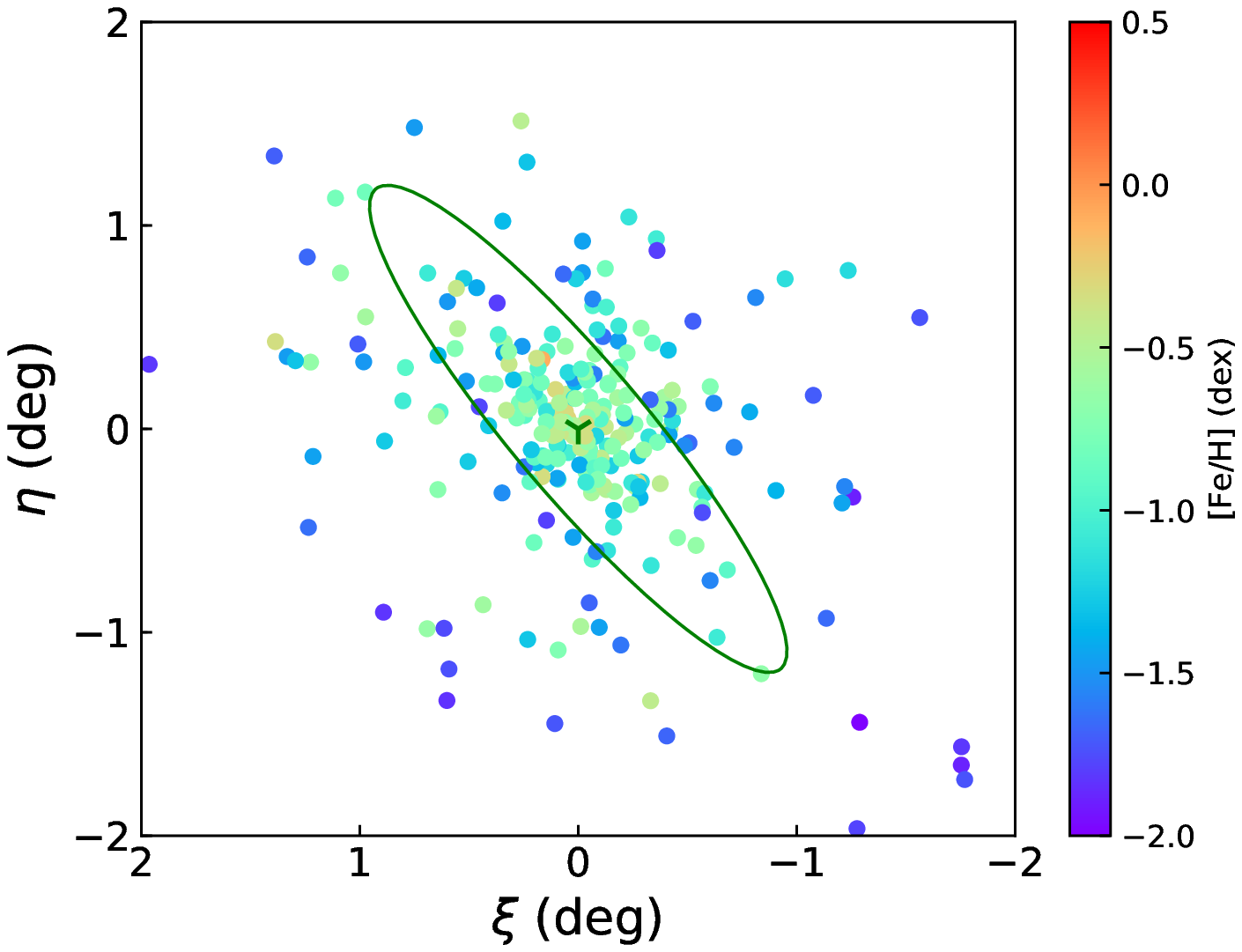}
}
\caption{
Spatial distributions of our catalogued star clusters. The coordinate (0,0) represents the centre of M31 (green `Y'; RA=$00^{\rm h}42^{\rm m}44^{\rm s}.30$, Dec=$+41\degr16^{\prime}09^{\prime\prime}.0$).  The green ellipse represents the optical disk of M31, for which we adopt a major axis length of 1.59\degr, a disk inclination angle of 78\degr, and a major axis position angle of 38\degr. The top panel shows the global spatial distribution of all the 346 sample clusters. Red and blue circles are respectively young and old clusters. The middle and bottom panels show the spatial distributions of young and old clusters, respectively, in the zoomed-in areas of the M31 disk. The colours represent the cluster metallicities.
}
\label{fig:coord}
\end{figure}

In total, we   selected 346 unique sources for our final catalogue. Some of them have been observed by LAMOST more than once. In total, 481 LAMOST spectra were collected for the catalogued objects. For clusters that have been observed  multiple times, we scale the spectra of lower $g$-band S/N by third-order polynomials to match the continuum level of the spectrum of the highest $g$-band S/N and then combine all the spectra together with weights of the inverse square of errors. We show the spatial distribution of all catalogued clusters of different ages and metallicities (see Section~4) in Fig.~\ref{fig:coord}. The sample spreads from the inner bulge to the outer halo of M31. The furthest cluster is located at projected radii $R\sim200\,\rm kpc$ from the M31 centre. At the region of $R = 25\sim150\,\rm kpc$, the cluster radial density profile exhibits a power-law decline \citep{2019MNRAS.484.1756M}.

\begin{table*}
\caption{Description of our M31 star clusters catalogue.}
\label{tab:catalogue}
\begin{center}
\begin{tabular}{ccl}
  \hline
  \hline
Column & Name & Description  \\
\hline
1 & name & name of the star cluster in the catalogue  \\
2 & RA &  Right Ascension (J2000)  \\
3 & Dec  & Declination (J2000) \\
4 - 16 &  c, d, e, ..., o and p  &  BATC $c,~d,~e,$ ... $o$ and $p$ magnitudes  \\
17 - 29 &  cerr, derr, eerr, ..., oerr and perr  &  BATC $c,~d,~e,$ ... $o$ and $p$ magnitude errors   \\
30 - 34 & su, sg, sr, si and sz & SDSS $u,~g,~r,~i$ and $z$ magnitudes \\
35 - 39 & suerr, sgerr, srerr, sierr and szerr & SDSS $u,~g,~r,~i$ and $z$ magnitude errors \\
40 & ebv & dust reddening $E(B-V)$ \\
41 & age & resultant age (log\,$t$) from this work \\
42 & ageerr & uncertainty of the age \\
43 & feh & resultant metallicity ([Fe/H]) from this work \\
44 & feherr & uncertainty of the metallicity \\
45 & RFclass & classification of clusters from the random forest classifiers \\
46 & class & final classification of clusters in the current work \\
\hline
\end{tabular}
\end{center}
\end{table*}

\subsection{Photometric data}

In addition to the LAMOST spectra, we  also collected multi-band photometries of the catalogued clusters to estimate their parameters. In the current work, the photometric data are  from two previous works.

 \citet{2015AJ....149...56M}  present multi-band photometry in 15 intermediate-band filters (Beijing-Arizona-Taiwan-Connecticut, BATC $a, b, c, ... o$ and $p$) of 304 globular clusters in M31 based on the observations from the BATC survey. The 15 intermediate-band filters cover the wavelength ranges of $3000\sim10000\,\rm \AA$. Of these clusters, 174 are in common with our catalogue. In the current work we use only the BATC $c$ to $p$ magnitudes of these catalogued clusters due to the large uncertainties of the BATC $a$ and $b$ magnitudes.

 \citet{2010MNRAS.402..803P} provide $ugriz$ photometry of 1595 M31 clusters and candidates based on images from SDSS. There are 301 objects with available photometric data in common with our catalogue. Seventeen objects are not found in the Peacock et al. catalogue, but have detections in SDSS. For these objects we adopt their Petrosian magnitudes from the SDSS archive.

As a result, there are 323 objects in our catalogue having photometric measurements in at least one of the BATC and SDSS filters. The reddening values of the individual clusters are selected from \citet{2012ApJS..199...37K} and \citet{2014AJ....148....4W}. There are 47 objects having no reddening values in the literature. For star clusters falling at a galactocentric distance larger than 22\,kpc we adopt a foreground reddening value of $E(B-V)$ = 0.13\,mag. For clusters at galactocentric distances smaller than 22\,kpc, the median reddening values of clusters within 2\,kpc radii of the clusters of concern are adopted. The magnitudes are then corrected with these reddening values by assuming a \citet{1989ApJ...345..245C} extinction law.

\begin{figure*}
\center{
\includegraphics[scale=0.5]{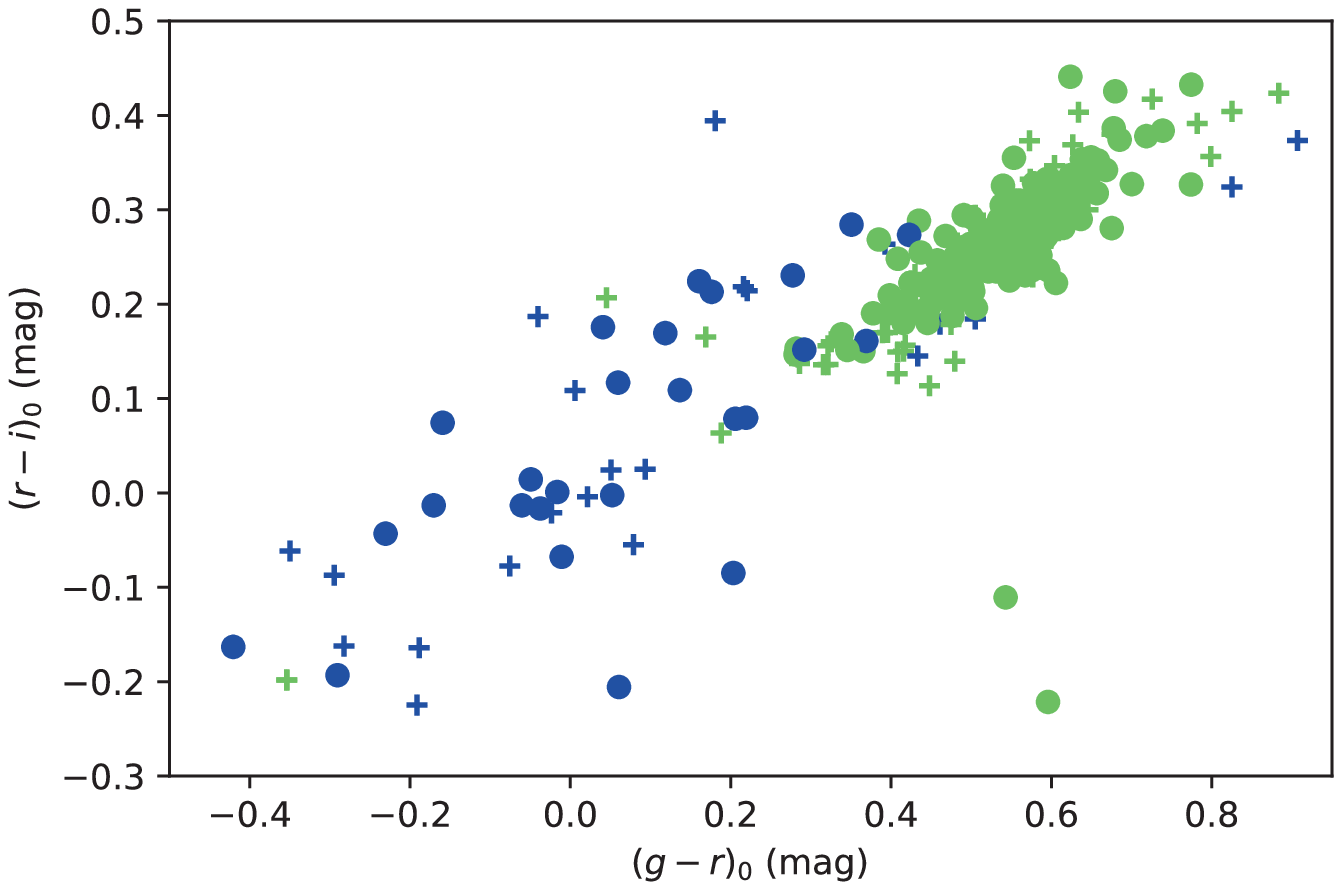}
\includegraphics[scale=0.5]{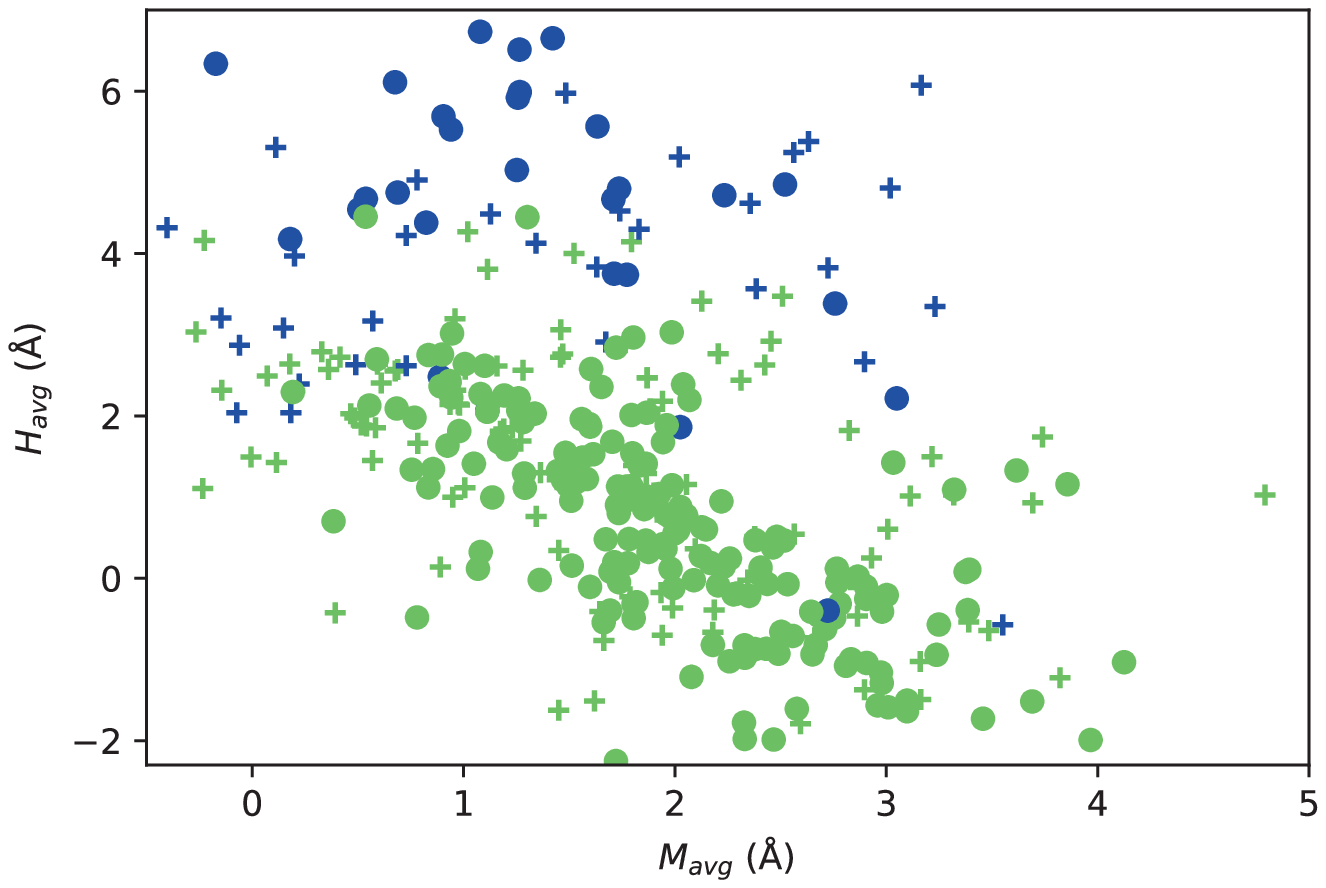}
}
\caption{Colour-colour (left panel) and line index-index diagrams of star clusters in M31. The indices are defined as $H_{\rm avg}$ = (H$\delta$F + H$\gamma$F + H$\beta$) $/ 3$ and $M_{\rm avg}$ = (Fe5270 + Mgb) $/2$. Blue and green pluses are respectively young and old clusters classified by the random forest classifiers. Blue and green circles are the young and old clusters in the training sample. }
\label{fig:svm}
\end{figure*}

\section{Determination of ages and metallicities}

\citet{2016AJ....152...45C}  estimate metallicities and ages of M31 clusters by the full spectrum fitting method, which fits the observed LAMOST spectra of the individual clusters pixel-by-pixel with the SSP model spectra. As indicated by \citet{2016AJ....152...45C} and \citet{2020arXiv200905712F}, there is a degeneracy between the young metal-rich clusters and the old metal-poor clusters for the full spectrum fitting method. In addition, for young massive clusters with ages   less than 1 - 2\,Gyr, the full spectral fitting algorithm may have significantly underestimated their metallicities but overestimated their ages. \citet{2016AJ....152...45C}  adopt the ages estimated by the SED fitting of multi-band photometry for young clusters. The SED fitting method is sensitive to the general shape of the continuum of the clusters over a broad wavelength range. The metallicities from the SED fitting method have large uncertainties due to the lack of information of the spectra lines  (e.g. \citealt{2010ApJ...725..200F,2010PASP..122.1164M}). There is also age--metallicity degeneracy for the SED fitting method if there is only optical photometry available (\citealt{1994ApJS...94..687W,1996stgi.conf.....L,2007MNRAS.381L..74K}). However, as indicated by \citet{2004MNRAS.347..196A},  we can break the age--metallicity degeneracy by including near-infrared photometry. Finally, the SEDs of old clusters ($>$ 1\,Gyr) have very similar shapes and they are usually hard to distinguish among themselves, thus the SED fitting method is usually only applicable to estimate ages of young clusters. To obtain more robust parameters of our catalogued M31 clusters, we have developed a new algorithm, presented  here, to include the information of both the spectra and the SEDs of the clusters. In the current work we will distinguish young clusters from old ones and then estimate the parameters separately for the two groups.

A random forest classifier is employed to classify the young and old clusters in our catalogue. The random forest classifier is a meta estimator that fits a number of decision tree classifiers on those sub-samples of the data set. It uses averaging to improve the predictive accuracy, and it controls overfitting. We build two random forest models, one with input parameters from the combination of the spectroscopic and photometric data and the other from the spectroscopic data only. For the first classifier, the input parameters are Lick indices from the LAMOST spectra and multi-band magnitudes from the SDSS photometry. Only Lick indices are used for the second classifier. We shift the LAMOST spectra to the rest frame using the radial velocities calculated by the ULySS code \citep{2009A&A...501.1269K}, smooth the spectra to match the Lick/IDS resolution, and then calculate the equivalent widths (EWs) of different lines by adopting the passbands defined in the literature (\citealt{1994ApJS...94..687W,1997ApJS..111..377W}). In the current work we adopt five line indices, including Mgb, Fe5270, H$\delta$, H$\gamma$, and H$\beta$. These lines are good indicators of the cluster metallicities and ages. They are the most significant lines in the LAMOST spectra of our sample clusters; they suffer smaller uncertainties compared to the other lines. For the SDSS photometry, we adopt all the $ugriz$ magnitudes. We created an empirical training set using catalogues from \citet{2009AJ....137...94C} and \citet{2011AJ....141...61C}. Based on spectra observed by the Hectospec fibre positioner and spectrograph on the 6.5m MMT, \citet{2009AJ....137...94C} and \citet{2011AJ....141...61C}   estimated the  parameters of $\sim$ 100 young and 300 old clusters in M31, respectively. We cross-matched our catalogue with those from \citet{2009AJ....137...94C} and \citet{2011AJ....141...61C}. Clusters with S/N in 4650\AA\ higher than 20 and having detections in all   five SDSS $ugriz$ bands are adopted as our training data set. This yields 28 young and 177 old clusters. The SCIKIT-LEARN package for PYTHON (\citealt{2012arXiv:1201.0490}) is used to build the models. The number of trees is set to be n\_estimators = 20. A classification rate of 98\%\ is obtained for the model with input parameters of both the Lick indices and SDSS magnitudes, while a rate of 96\%\ is obtained for the model with Lick indices as input parameters only. We then apply the first model to our catalogued clusters with all the five SDSS magnitudes and the second model to clusters without. The result is shown in Fig.~\ref{fig:svm}. Our classification seems to work well for most of the clusters, except for a few outliers. Compared to the old clusters, young clusters have bluer colours and higher H-line indices. As a result, we have 63 and 283 clusters classified as young and old clusters by the random forest classifiers in our catalogue.

\subsection{Determining ages and metallicities  of young clusters}

We estimated the ages of the young clusters  by SED  fitting of multi-band photometry (demonstrated to be reliable
in \citealt{2018AJ....156..191F}). We fit the de-reddened data introduced in Sect.~2.3 with predictions of the SSP models of \citet{2003MNRAS.344.1000B} (hereafter BC03). The BC03 models with initial mass function (IMF) from \citet{2003PASP..115..763C} and the stellar library from the STELIB spectral library (\citealt{2003A&A...402..433L}) are adopted in the current work. The ages of the model ranges between  5.0 $<$ log\,$t$ $<$ 10.3\,(yr) with log\,$t$ bins varying from 0.005 to 0.05\,(yr). The model uses only six metallicities ($Z$ = 0.0001, 0.0004, 0.004, 0.008, 0.02, and 0.05). We linearly interpolate the model to an interval of [Fe/H] of 0.01\,dex. This yields 62,040 BC03 model spectra.

\begin{figure}
\center{
\includegraphics[scale=0.6]{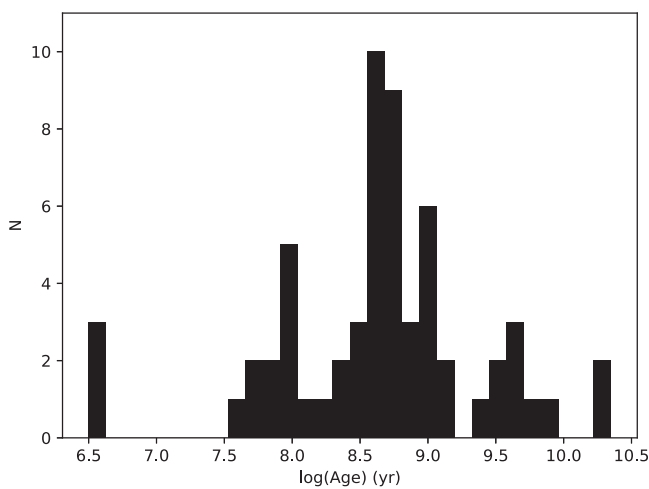}
}
\caption{Distribution of SED ages of the sample objects classified as young clusters from the random forest classifiers.}
\label{fig:sedage}
\end{figure}

We applied SED fitting to all the catalogued star clusters that are detected in at least four bands in SDSS or ten bands in BATC. The cluster ages are then determined by the least-squares method. As a result, we  obtain SED ages of 60 clusters in our sample that are classfied as young objects by the random forest classifiers. There are three clusters classified as young clusters by the random forest model having no SED ages due to the lack of photometric data. We do not provide ages of these three clusters as the ages derived from the spectral fitting would be overestimated \citep{2016AJ....152...45C}. The distribution of the SED ages is shown in Fig.~\ref{fig:sedage}. Most of these clusters have best-fit SED ages smaller than 1.6\,Gyr. There are ten clusters having SED ages larger than 2\,Gyr. We   checked these ten clusters in the colour-colour and line index-index diagrams; they should be old clusters with red colours. However, their line indices are located at the region where there are both young and old clusters ($H_{\rm avg}$ $\sim$ 2 - 4). We suggest that these clusters are misclassified as young clusters by the random forest classifiers. In the current work these ten sources are re-classified as old clusters.

We then derive the metallicities of these young clusters by fitting the principle components of their LAMOST spectra with those from the BC03 model. We shift the LAMOST spectra to the rest frame using the radial velocities calculated by ULySS code. Then we degrade the spectral resolution of both the LAMOST and the BC03 model spectra to $\Delta \lambda$ = 5\,\AA. Both the observed and model spectra are normalized to unity by a third-order polynomial. Principal components are extracted from the matrix composed by the observed and model spectra by the principal components analysis (PCA). In this work we   adopt ten components for each spectrum. As the young clusters are expected to be metal rich, we fit the observed spectra principle components of the young clusters only to those of the model with metallicities [Fe/H] ranging from $-$1.0 to 0.5\,dex. For each source we select the BC03 models of the age closest to its SED age. The metallicity is then determined by the least-squares method.

\subsection{Determining ages and metallicities  of old clusters}

We construct a  method similar to that of  \citealt{2020ApJS..246....9Z} to estimate the parameters of old clusters. For the 293 old clusters in our catalogue we adopt the \citet{2010MNRAS.404.1639V} models that are based on the empirical stellar spectral library of medium resolution INT library of empirical spectra (MILES; \citealt{2006MNRAS.371..703S,2007MNRAS.374..664C}). Spectra from the models cover the optical wavelength range of $3540.5\sim7409.6\,\rm\AA$ with a nominal resolution of full width at half-maximum (FWHM) of $2.3\,\rm \AA$. The models are based on the solar-scaled theoretical isochrones of \citet{2000A&AS..141..371G} and several different IMFs. In this work we adopt the models calculated with the IMF from \citet{1955ApJ...121..161S}. The models cover ages of $10^{8}<t<1.5\times10^{10}$\,\rm yr and metallicities of $-2.32<\,\rm [Fe/H]<0.22\,\rm dex$. To obtain robust ages and metallicities of the catalogued old clusters, we  combine both the spectroscopic and photometric data.

In the current work we use the Vazdekis et al. models for the old clusters because the models are based on the MILES empirical spectral library which consists of spectra of real stars. The MILES spectra have a spectral resolution comparable to that of the LAMOST spectra, and are accurately flux-calibrated with an accuracy of a few per\,cent. However, due to the lack of spectra of young stars in the MILES library, the Vazdekis et al. SSP models only extend to an age of 10$^8$\,yr. We thus adopt the BC03 models for the young clusters. We  compared the ages and metallicities derived using respectively the Vazdekis et al. and BC03 models for our catalogued young clusters of ages between $10^{8}$ and $1.5\times10^{10}$\,yr. We find that compared to the resultant parameters derived from the BC03 models, the ages derived from the Vazdekis et al. models are overestimated and the metallicities underestimated (see also \citealt{2016AJ....152...45C}). For old clusters we  also adopted other SSP models, such as the PEGASE-HR models \citep{2004A&A...425..881L}. The results are similar to those derived from the Vazdekis et al. models.

Due to the limited wavelength range of the Vazdekis et al. models ($3540\sim7410\,\rm\AA$), we use only the photometric magnitudes of clusters in the BATC $c$ to $j$ bands and SDSS $g$ and $r$ bands. In this work we  use the adjacent colours of clusters:  BATC $c-d, d-e, e-f, f-g, g-h, h-i, i-j$, and SDSS $g-r$. As the LAMOST spectra contain thousands of pixels while there are only eight multi-band colours, in this work we adopt the principle components of the spectra instead. Again, we shift the LAMOST spectra to the rest frame using the radial velocities calculated by ULySS. The spectral resolution of the Vazdekis et al. model spectra are degraded to the same resolution as the LAMOST. As the Vazdekis et al. models have only seven metallicities, we linearly interpolate the models to an interval of [Fe/H] of 0.01\,dex. This yields 12,495 SSP model spectra. Both the observed and model spectra are normalized to unity by a third-order polynomial. Principal components are extracted from the matrix composed of the observed and model spectra by the PCA analysis, and we adopt ten components for each spectrum.

The 12,495 interpolated Vazdekis et al. models are randomly divided into two sub-samples, a training sample consisting of 80\% of the models and a test sample containing the remaining 20\% models. We use the training sample to generate the random forest models and the test sample to validate the generated relations. Based on the training sample, we construct non-parameter relations between the observed data (spectral principal components and/or colours) D and the parameters of clusters (metallicity [Fe/H] and age $t$) as
\begin{equation}
\begin{aligned}
D_{i}=f_{i}([\rm Fe/\rm H], \it{t}),
\end{aligned}
\end{equation}
where $i$ is the index of the principal components and colours, $D_{i}$ the $i$th spectral principal component or colour, and $f_{i}$ the $i$th random forest model. In the current work we have ten spectral principal components and eight colours (i.e. $i=1,2,3,...,18$). We  obtain the above relations ($f_{i}$) via the python wrapper of ENSEMBLE in the scikit-learn package \citep{2012arXiv:1201.0490}. We use ten estimators. Other parameters are set by default. Good agreements are found for all these relations (scores higher than 0.996) by applying the random forest models to the test sample clusters.

Based on the resultant random forest relations, we then calculate the parameters of our catalogued clusters by fitting their observed spectral principal components and colours. The Markov chain Monte Carlo (MCMC) method is adopted to find the best-fit parameters that maximize the likelihood
\begin{equation}
\begin{aligned}
\textup{lg}L=-\sum_{i}[D_{i}^{\textrm{obs}}-f_{i}([\textrm{Fe}/\textrm{H}],t)]^{2},
\end{aligned}
\end{equation}
where $D_{i}^{\rm obs}$ and $f_{i}([\textrm{Fe/H}],t)$ are the $i$th observed spectra principal component and/or colour and the random forest model output with the given metallicity [Fe/H] and age $t$, respectively. The $\sc python$ code $emcee$ \citep{2010CAMCS...5...65G} is applied in the current work. The resulting parameters and their uncertainties are estimated in terms of the 50th, 16th, and 84th percentiles of the accepted samples from the MCMC chain.

\begin{figure*}
\center{
\includegraphics[scale=0.8]{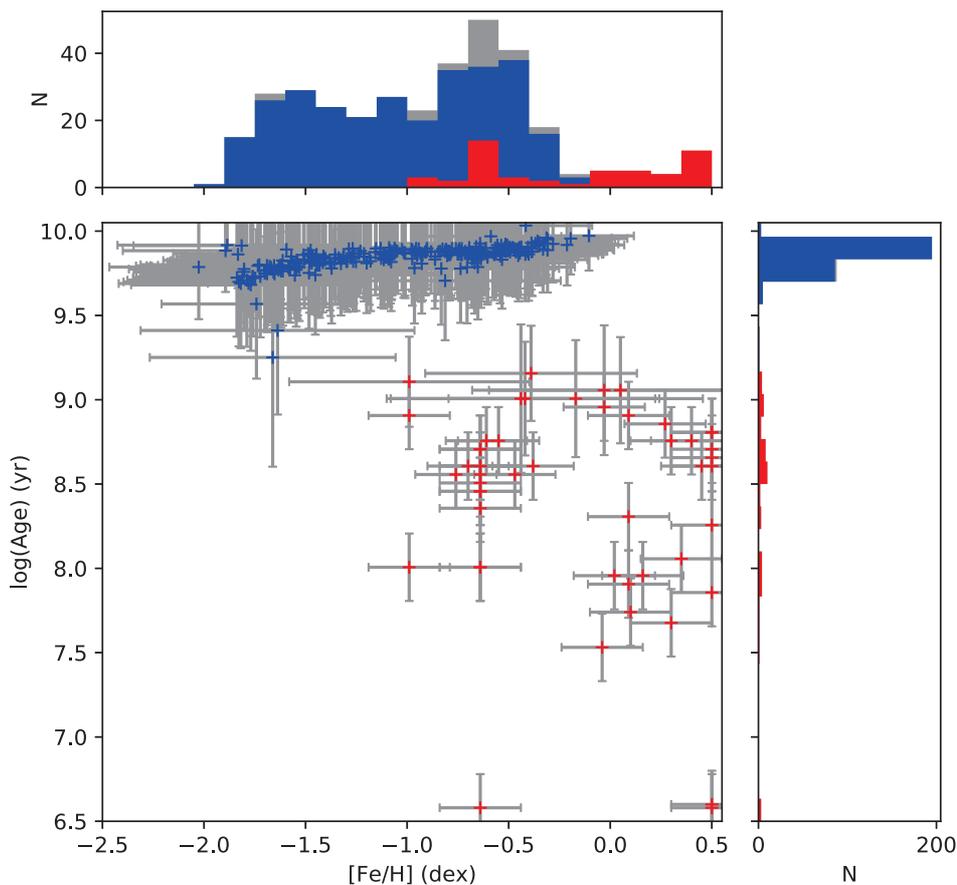}
}
\caption{Metallicities vs ages of our sample clusters derived in the current work. Red and blue pluses (with grey error bars) represent the catalogued young and old clusters, respectively. Histograms of age and metallicity distributions of clusters are respectively plotted on the right and at the top of the diagram. The red, blue, and black histograms give distributions of young, old, and all clusters, respectively.
}
\label{fig:agefeh}
\end{figure*}

\begin{figure}{}
\center{
\includegraphics[scale=0.6]{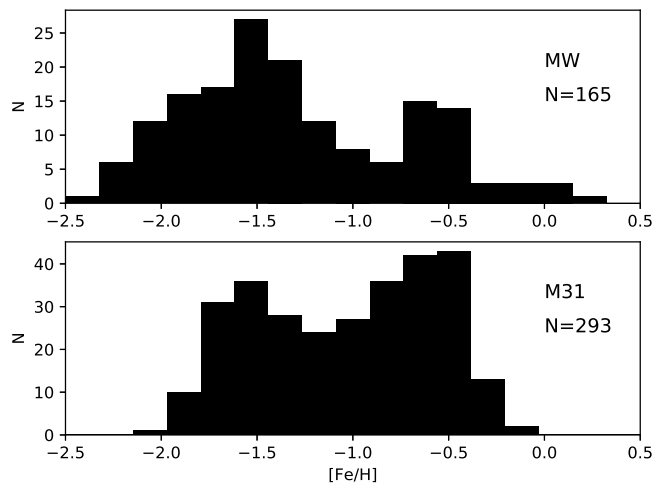}
}
\caption{Comparison of metallicity distributions for 165 Galactic globular clusters (upper panel) and 293 M31 old clusters (lower panel).}
\label{fig:distribution}
\end{figure}

\begin{figure*}
\center{
\includegraphics[scale=0.45]{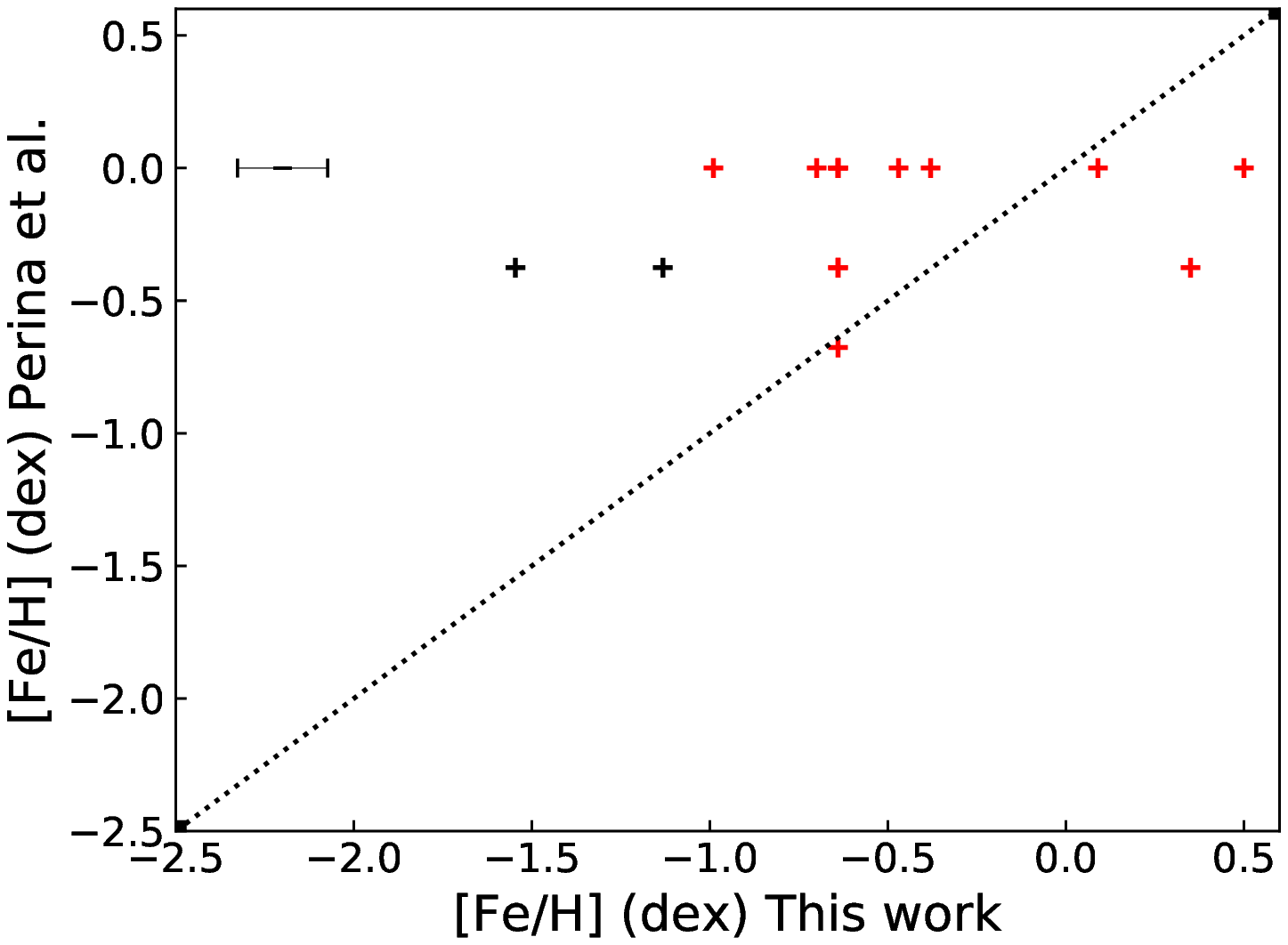}
\includegraphics[scale=0.45]{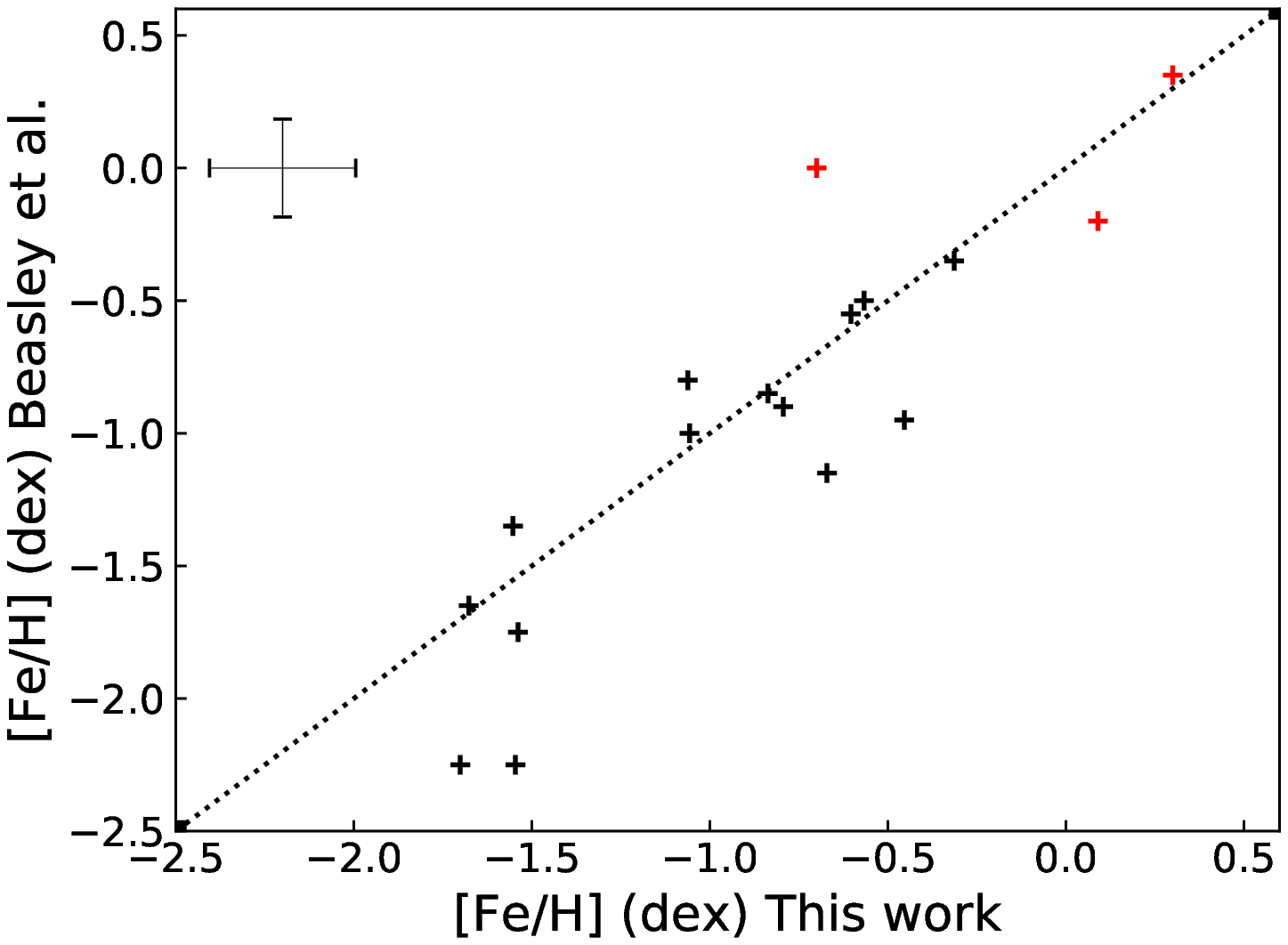}
\includegraphics[scale=0.45]{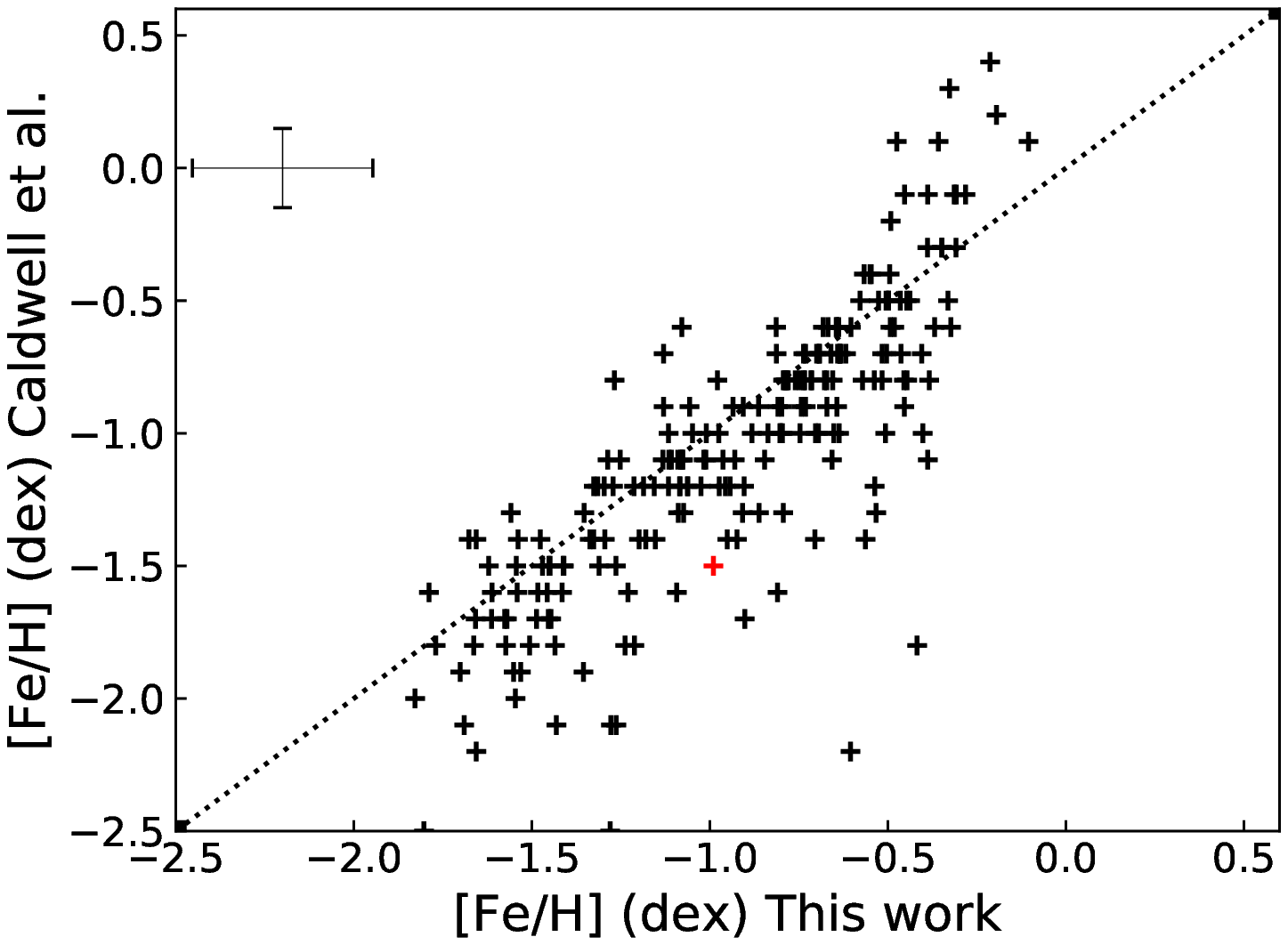}
\includegraphics[scale=0.45]{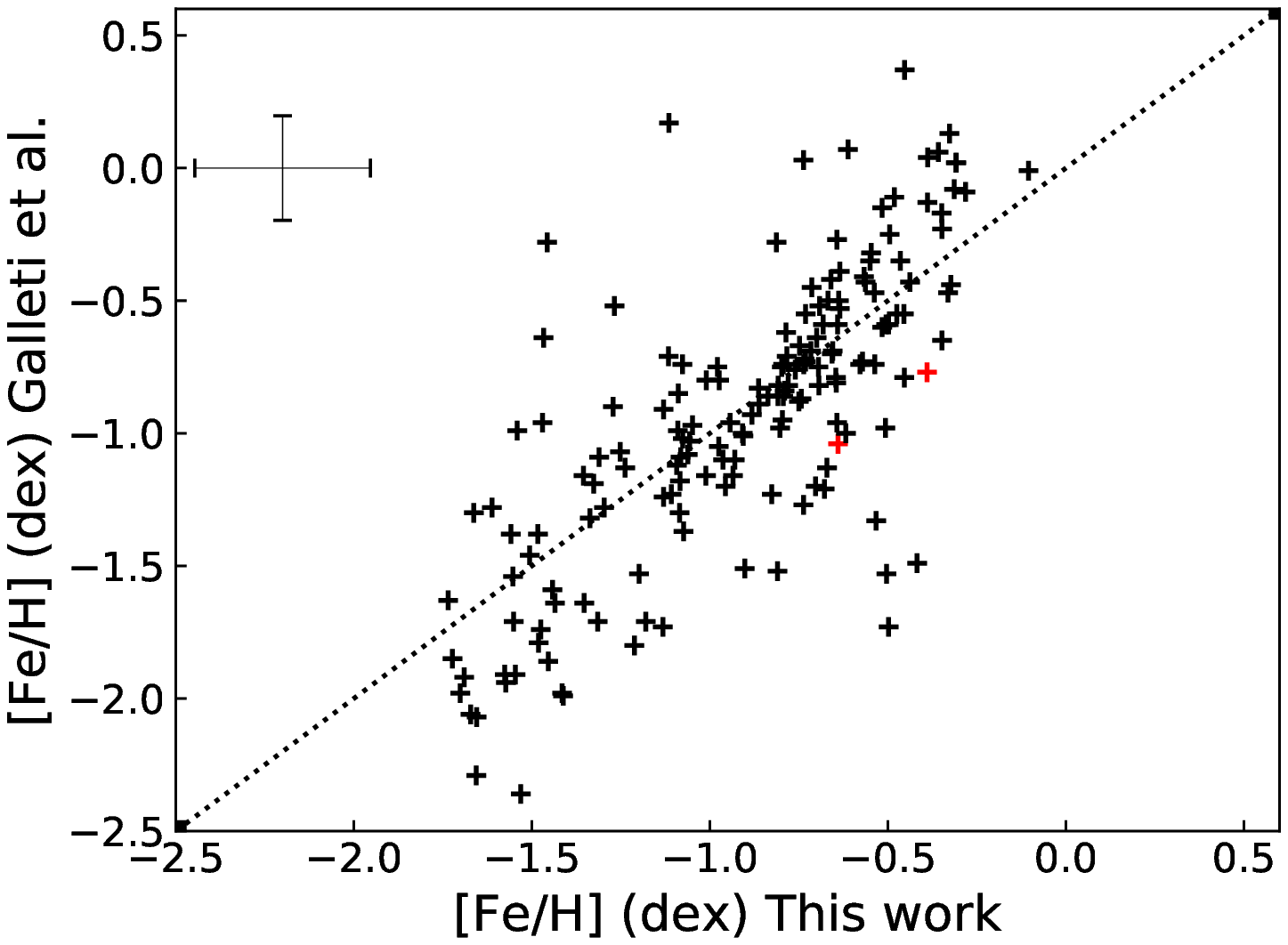}
\includegraphics[scale=0.45]{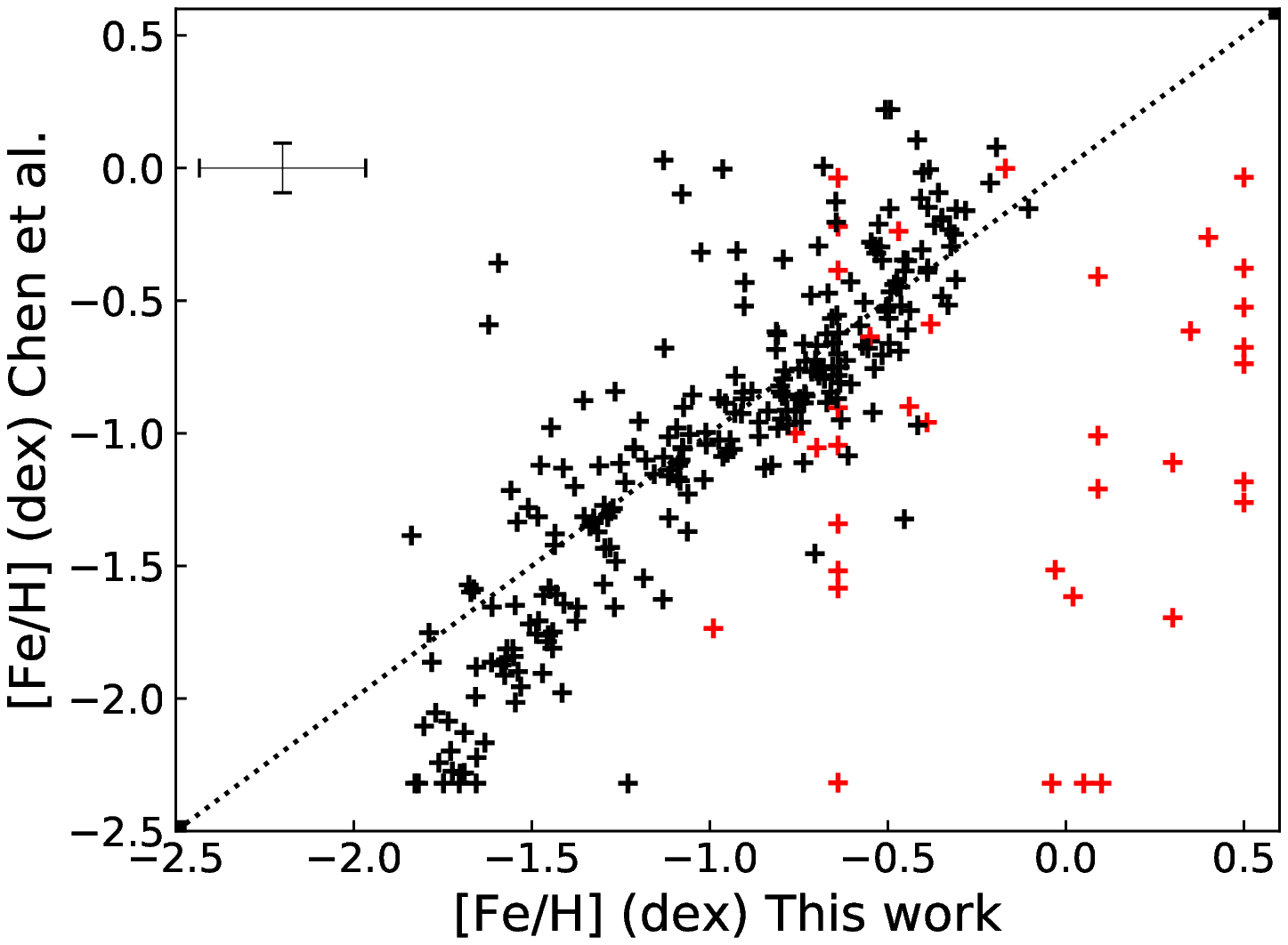}
\includegraphics[scale=0.45]{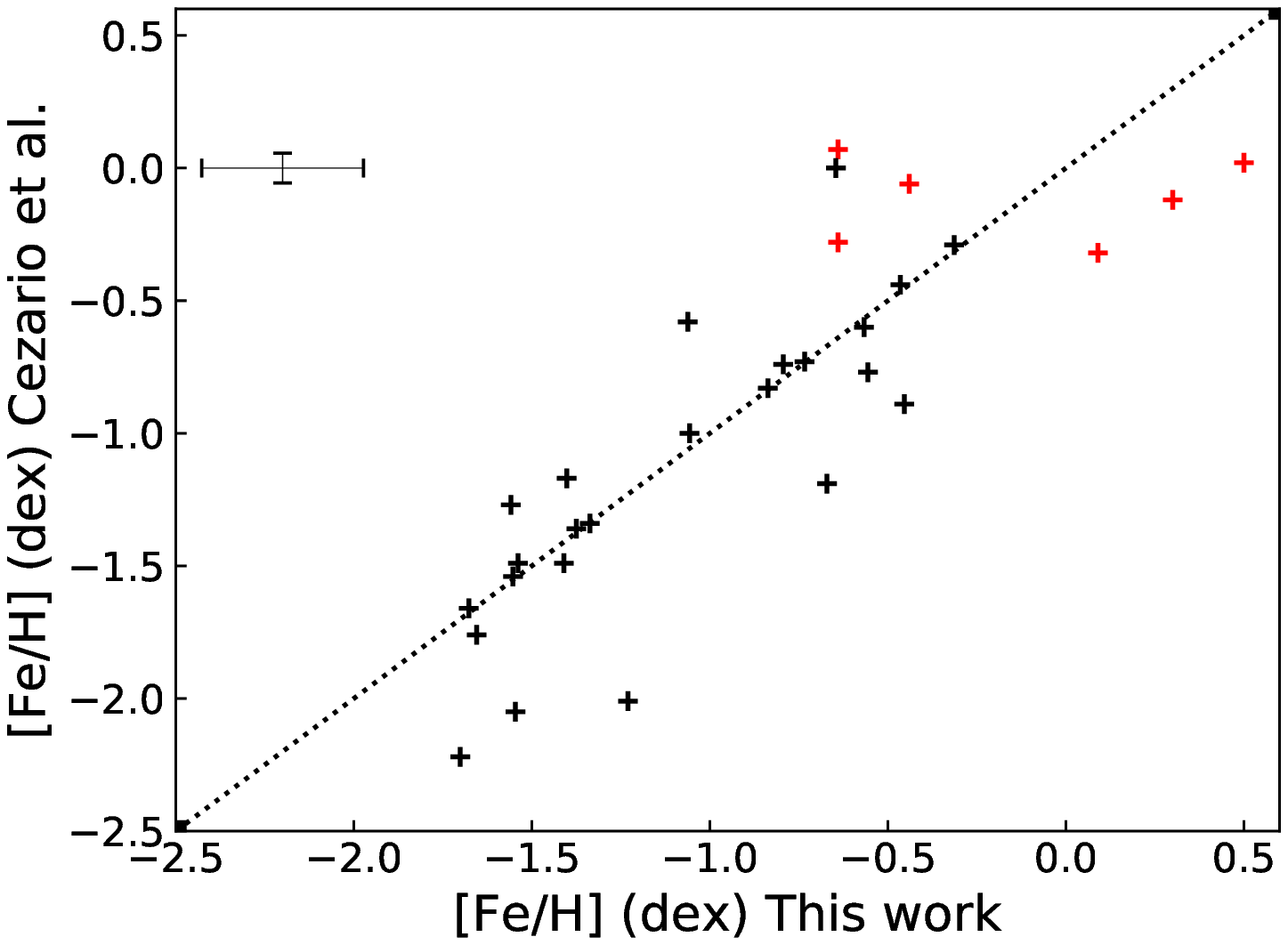}
\includegraphics[scale=0.45]{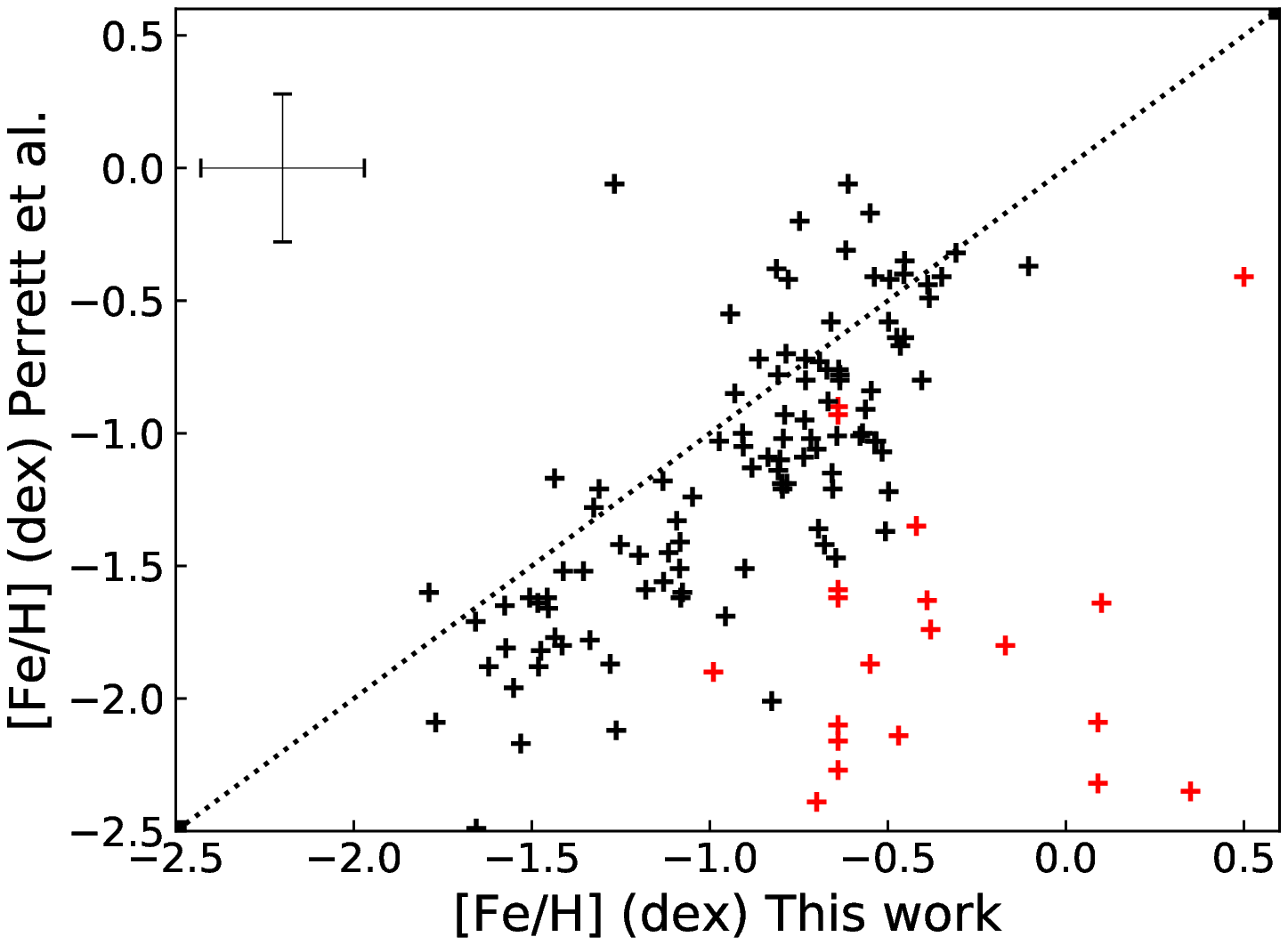}
\includegraphics[scale=0.45]{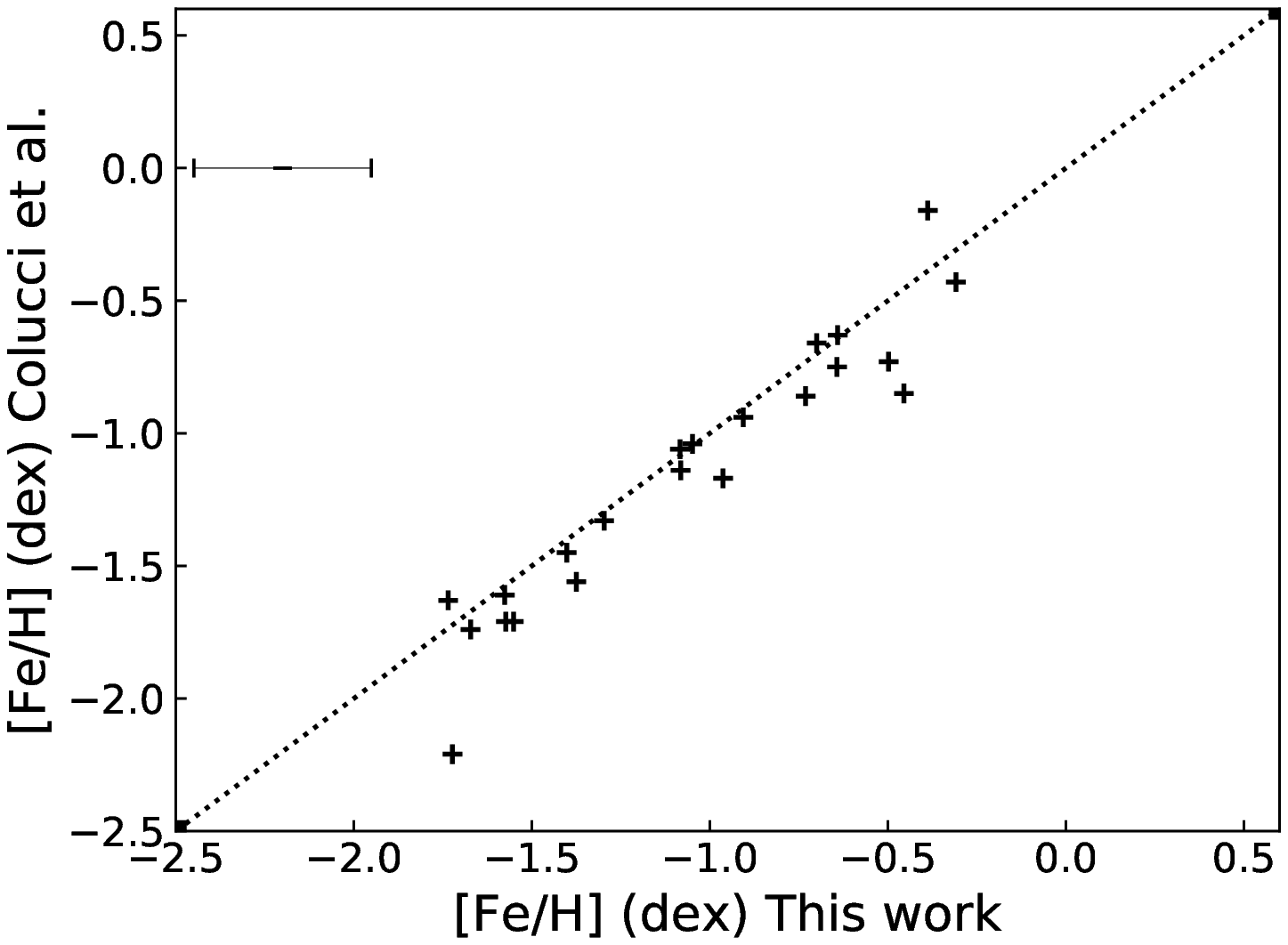}
}
\caption{Comparison between [Fe/H] derived in the current work and those from previous works. Our metallicities for the young and old clusters are plotted in red and black, respectively. The error bar in each panel shows the median metallicity errors.
}
\label{fig:feh}
\end{figure*}

\begin{figure*}
\center{
\includegraphics[scale=0.45]{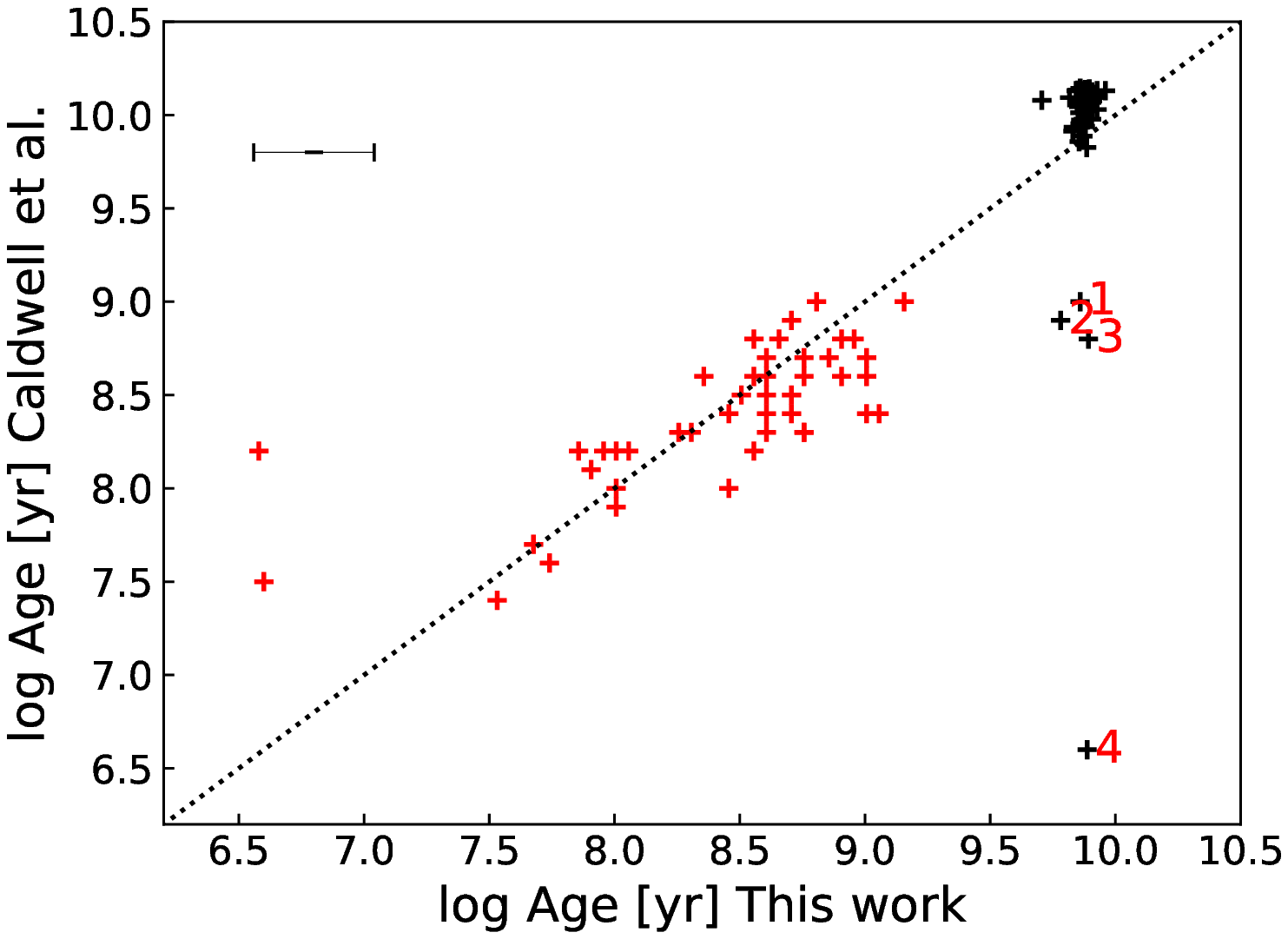}
\includegraphics[scale=0.45]{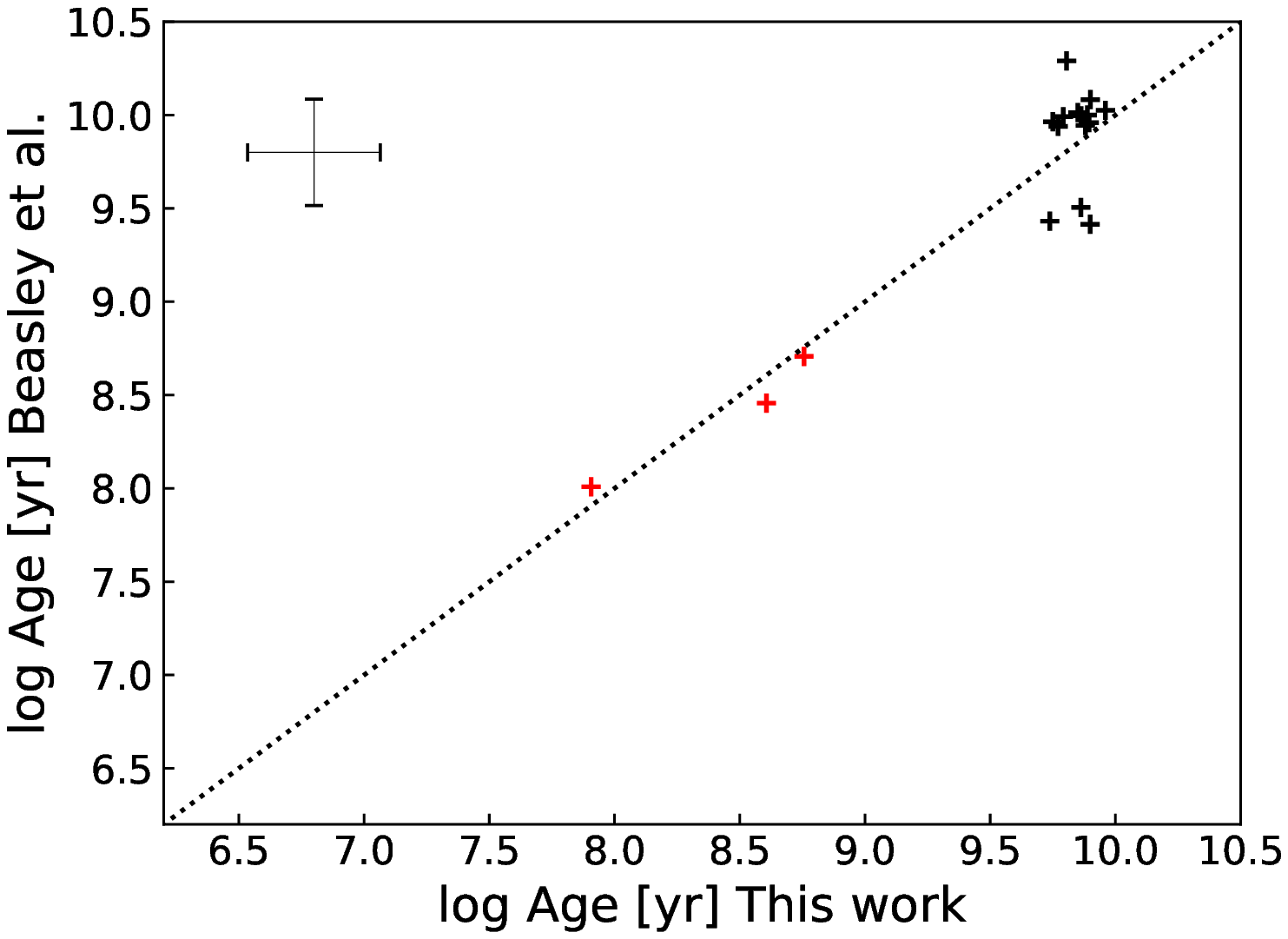}
\includegraphics[scale=0.45]{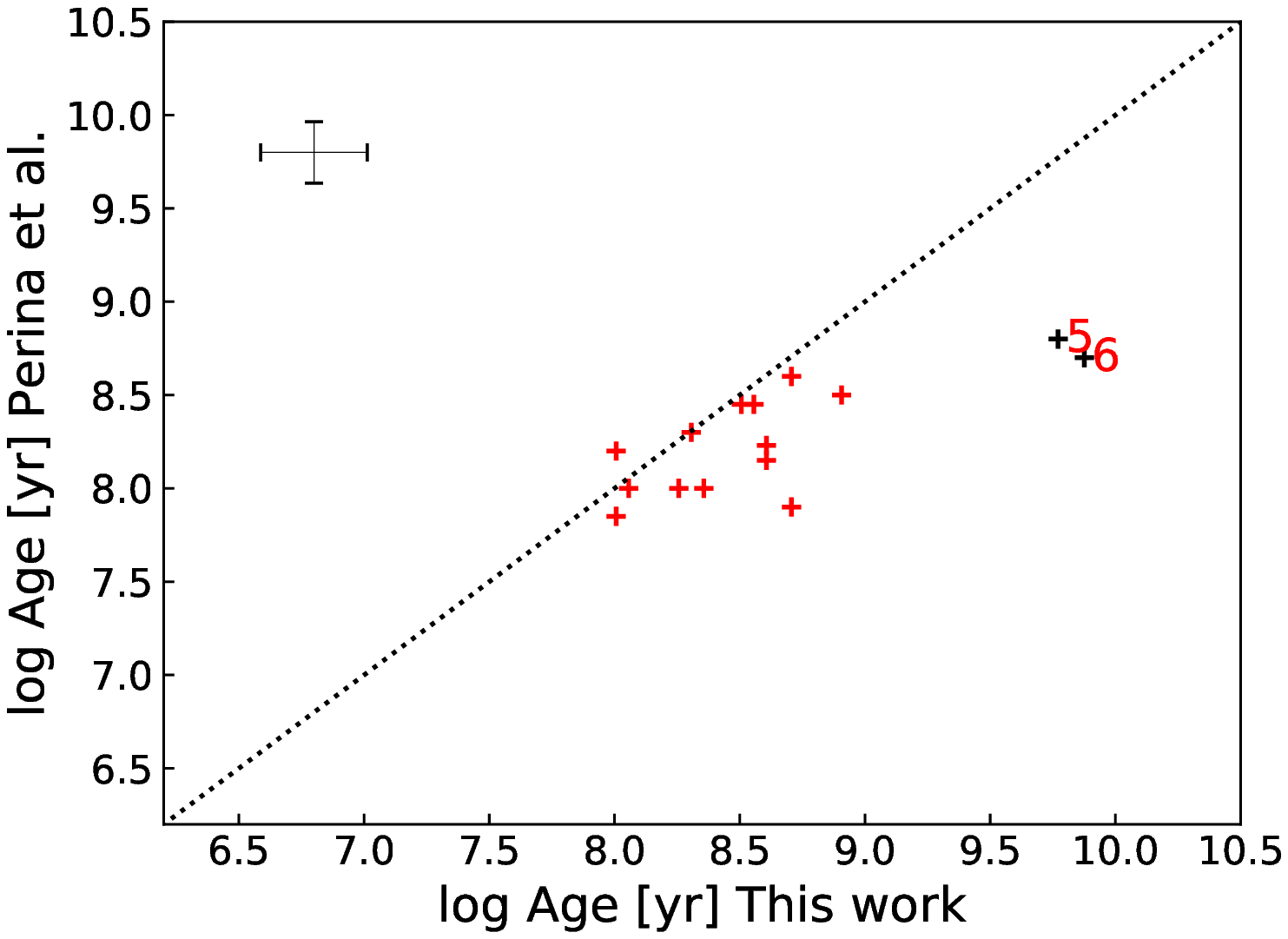}
\includegraphics[scale=0.45]{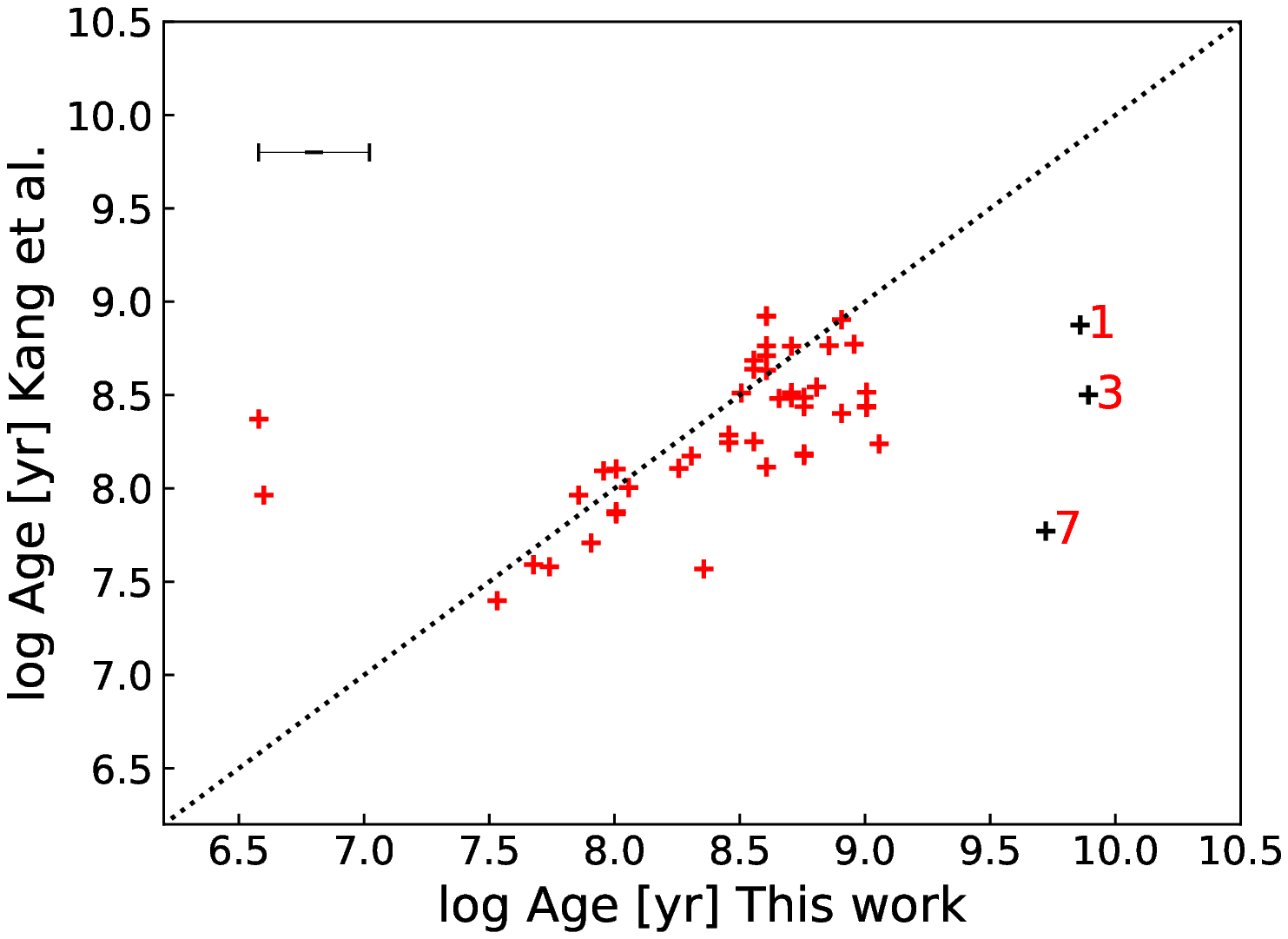}
\includegraphics[scale=0.45]{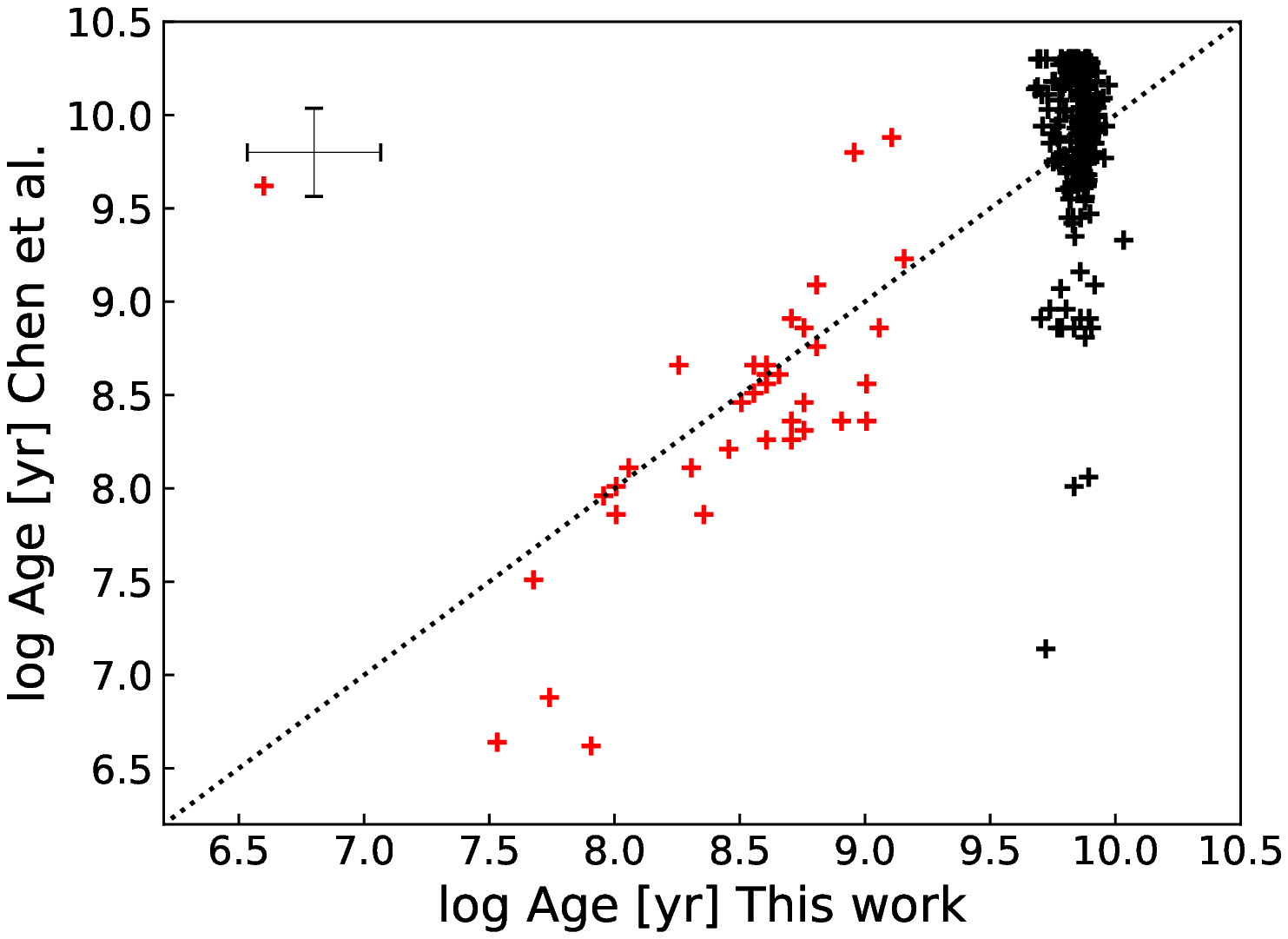}
\includegraphics[scale=0.45]{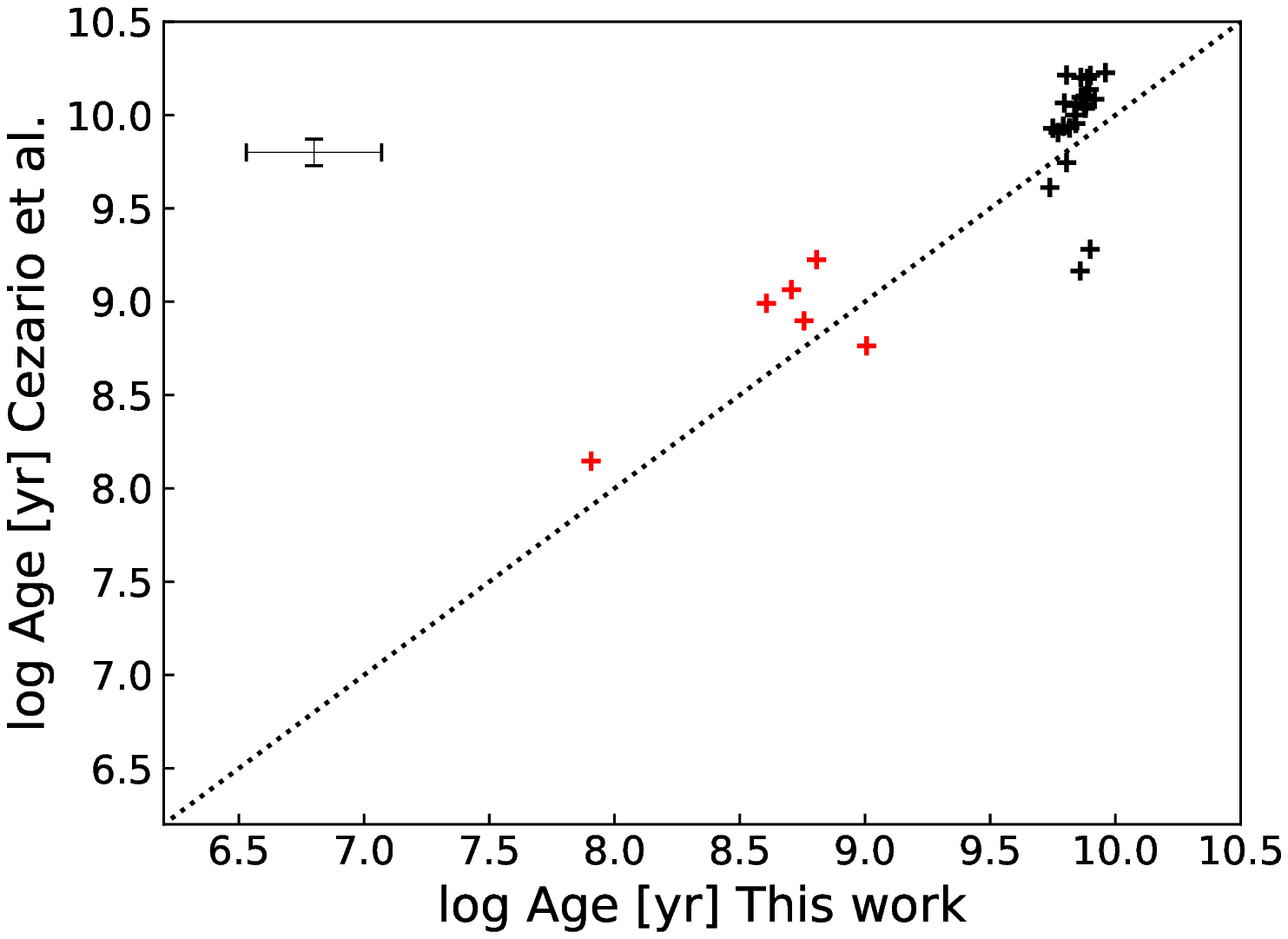}
\includegraphics[scale=0.45]{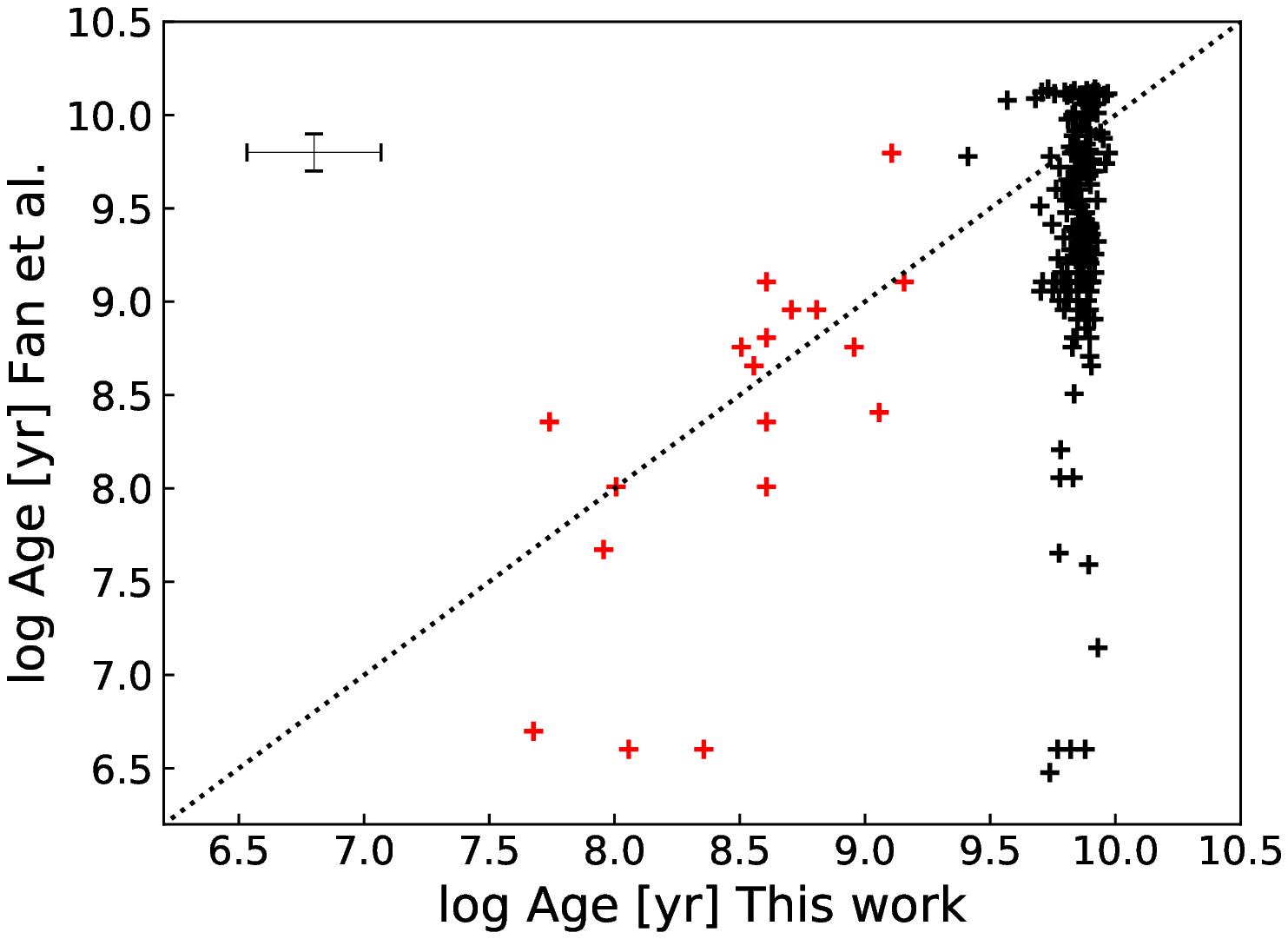}
\includegraphics[scale=0.45]{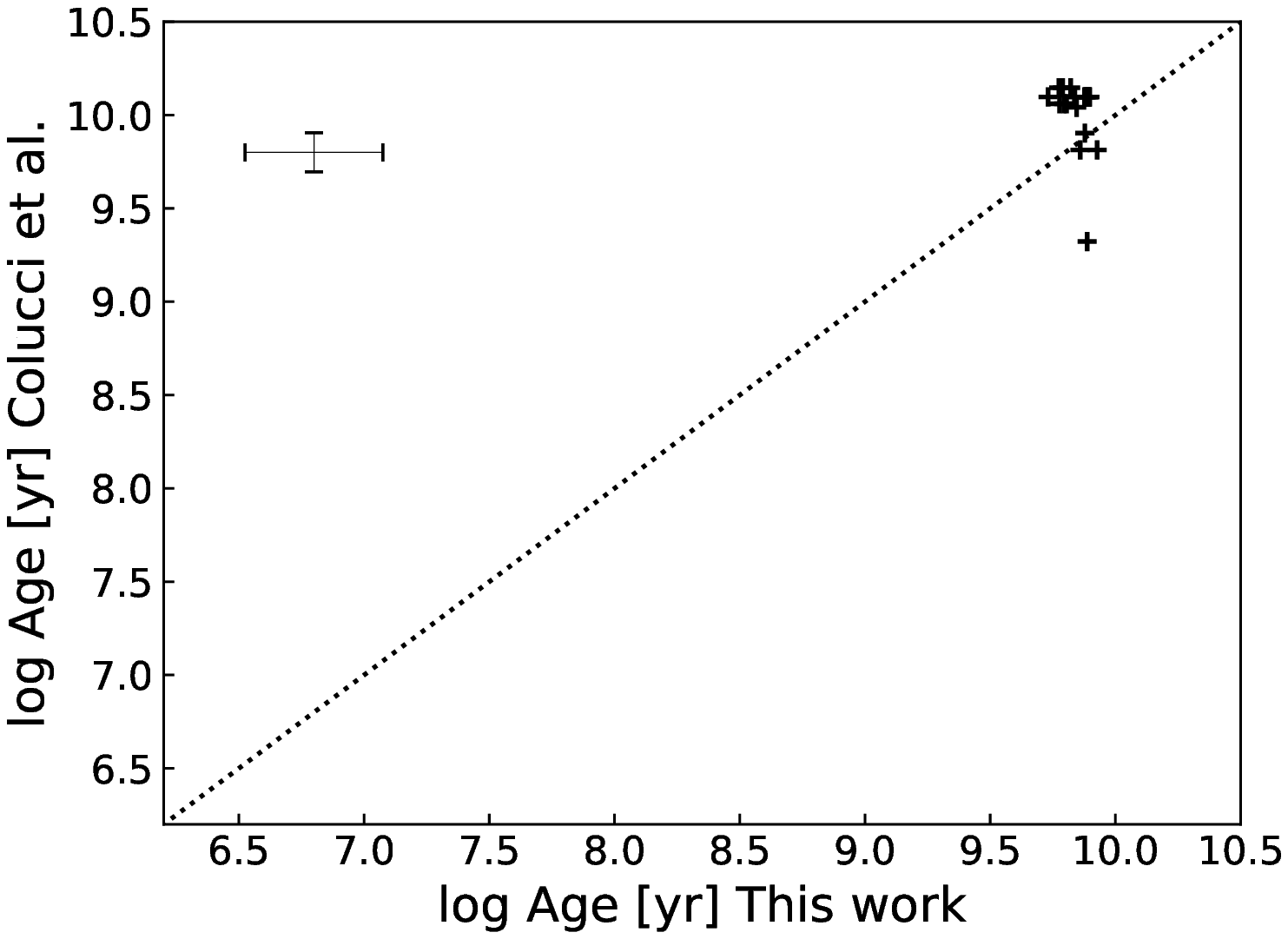}
}
\caption{Comparison between our derived ages and those from previous works. Our ages for the young and old clusters are plotted in red and black, respectively. The error bar in each panel shows the median age errors. The red numbers  $1-7$ are respectively the clusters B316-G040, B349, B089D, B240D-D066, B347-G154, B083-G146, and SK007A.
}
\label{fig:ages}
\end{figure*}

\begin{figure*}
\center{
\includegraphics[scale=0.45]{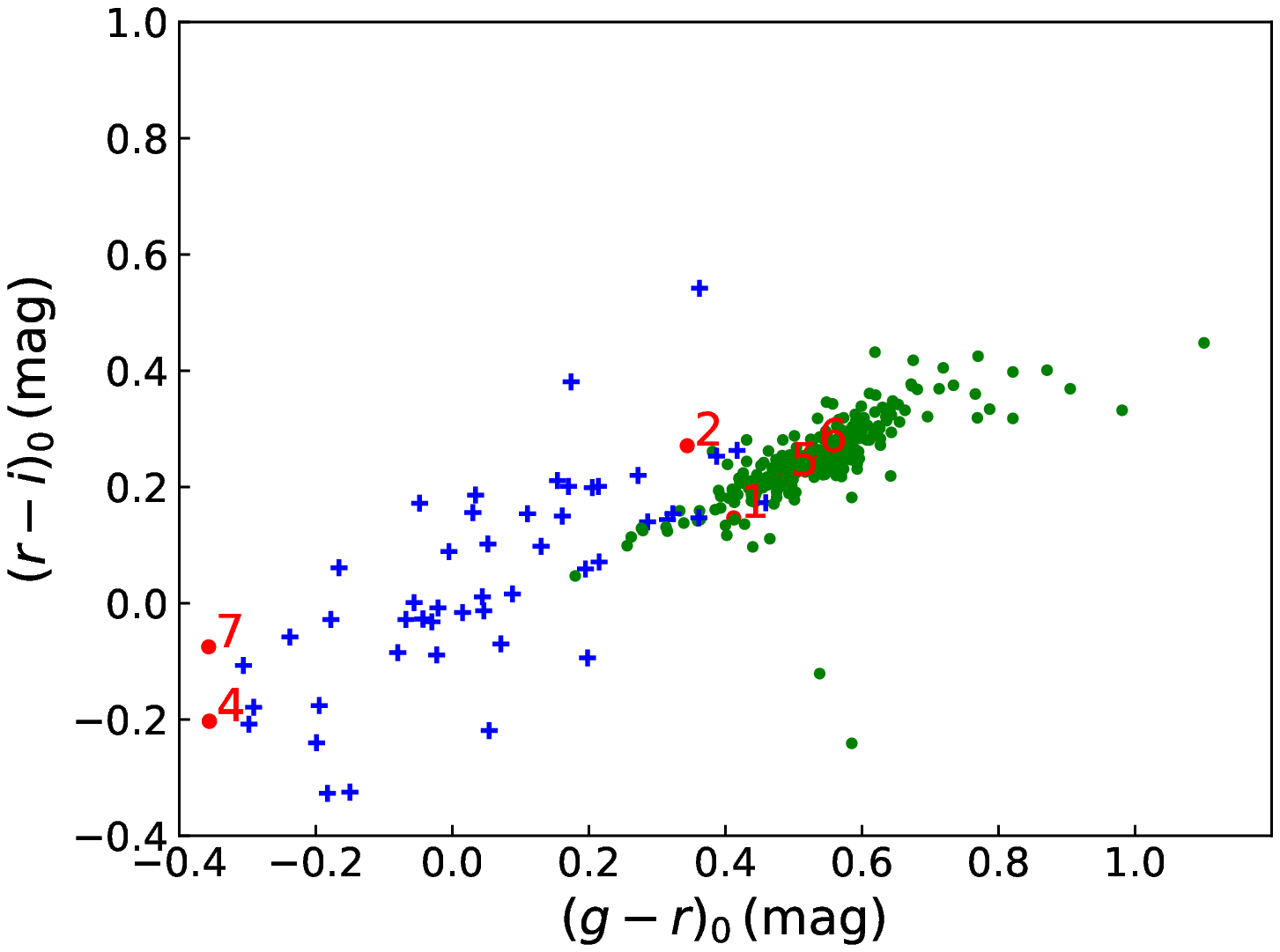}
\includegraphics[scale=0.45]{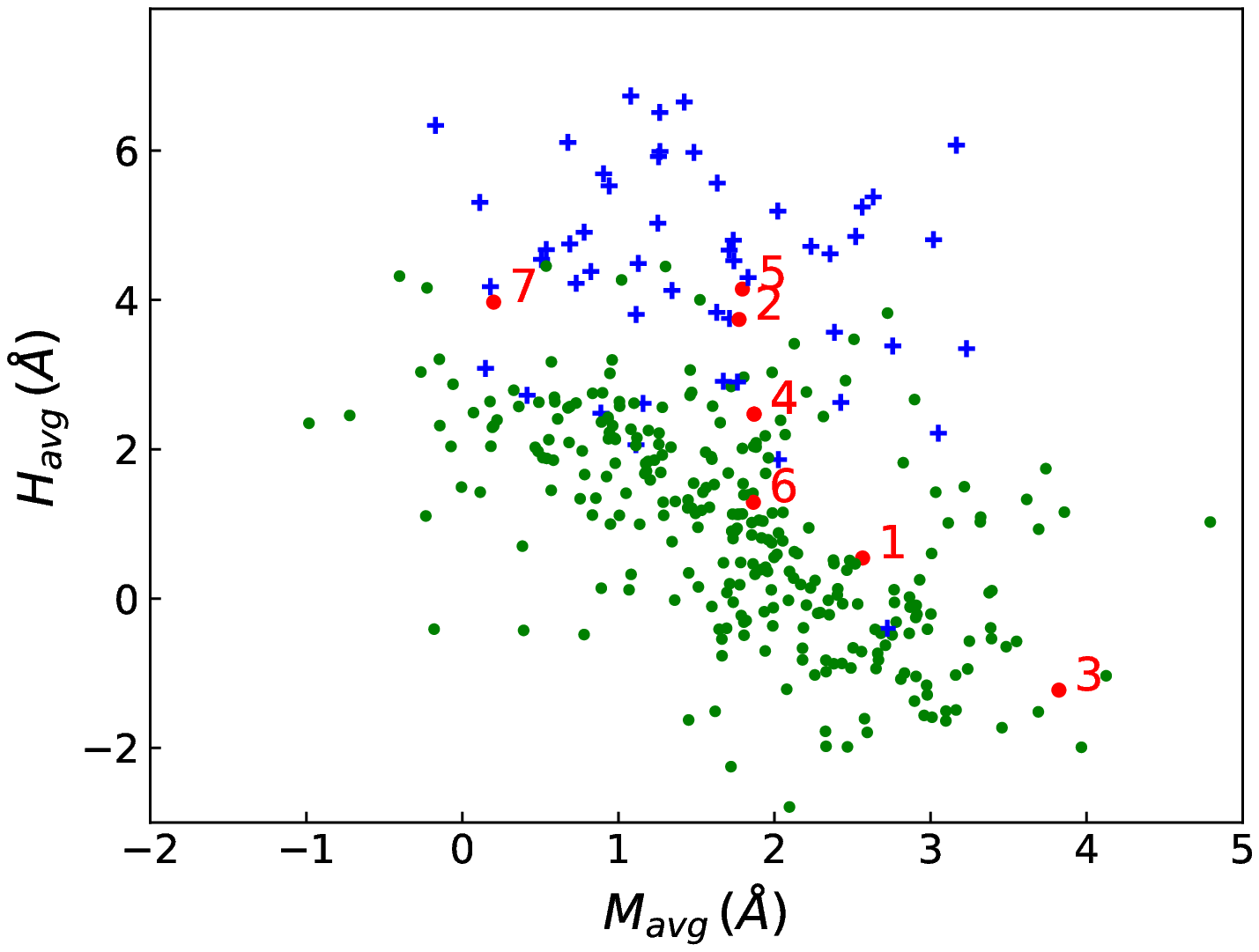}
}
\caption{Positions of seven clusters in the colour-colour and line index-index diagrams. Blue pluses and green circles correspond to young and old clusters in our catalogue, respectively. The red numbers  $1-7$  are respectively the clusters B316-G040, B349, B089D, B240D-D066, B347-G154, B083-G146, and SK007A.
}
\label{fig:outliers}
\end{figure*}

\section{Results}

A catalogue that contains the coordinates, multi-band photometries, and the resultant ages and metallicities of the 346 sample clusters is available in the online version of this manuscript\footnote{The catalogue is also available online at \url{http://paperdata.china-vo.org/Wang.SC/2020/AA/table1.fits}.}. The data format of the catalogue is described in Table~\ref{tab:catalogue}. Figure~\ref{fig:agefeh} displays the relationship between the yielded metallicities and the ages of our sample clusters. The gap between the old and young clusters is clearly shown in the figure, as proved in previous studies (\citealt{2009AJ....137...94C,2011AJ....141...61C,2016AJ....152...45C}). The young clusters have a broad metallicity distribution from $-$1\,dex to 0.5\,dex. The old clusters  have a metallicity distribution ranging from $-2.1$ to 0\,dex. Most of the old clusters have derived ages of around 10\,Gyr.

In Fig.~\ref{fig:distribution} we compare the metallicity distribution of our catalogued M31 old clusters with those of the Galactic globular clusters. The Galactic globular clusters are collected from  \citet{2006A&A...450..105B} and \citet{2009A&A...505..139C}. Visually the metallicity distribution of M31 old clusters has  two peaks at [Fe/H] $\sim$ $-$0.6 and $-$1.5\,dex. This is consistent with the findings of the previous works (\citealt{2009A&A...508.1285G,2011AJ....141...61C}). However, as there are many M31 clusters having [Fe/H]  at around $-$1\,dex, we do not think that the distribution is clearly bimodal, like the Milky Way distribution which has an obvious dip at [Fe/H] = $-$0.9\,dex. The primary peak of the M31 distribution occurs at [Fe/H]$\sim-$0.6\,dex, significantly more metal rich than in the Milky Way case  where the maximum is at [Fe/H]$\sim-$1.6\,dex. M31 also has   a much larger fraction of metal-rich old clusters with respect to the Milky Way.

In our catalogue the ages of $\sim$ 30 clusters and the metallicities of $\sim$ 40 clusters are derived for the first time. The remaining clusters have parameters determined in previous works by different methods. We then compare the parameters of these clusters with those from the literature. The comparisons are shown in Fig.~\ref{fig:feh} and Fig.~\ref{fig:ages}. In the top two panels of Fig.~\ref{fig:feh} we compare our derived metallicities to those from \citet{2009A&A...507.1375P} and \citet{2004AJ....128.1623B}. There are only a few works that estimate the metallicities of young clusters in M31. \citet{2009A&A...507.1375P} obtained ages and metallicities of a group of possible young massive clusters in M31 by fitting the colour-magnitude diagrams (CMDs) from the Hubble Space Telescope (HST) observations with the theoretical isochrones. \citet{2004AJ....128.1623B} derived metallicities of 30 star clusters in M31 based on their spectra. There is   larger scatter for the comparisons of metallicities of young clusters between our results and those from \citet{2009A&A...507.1375P} and \citet{2004AJ....128.1623B}. This is partly due to the large uncertainties found in  our data and in the  literature. However, our estimates are roughly consistent with the literature values except for a few outliers. We have also compared our estimated metallicities of old clusters to those analyzed by \citet{2004AJ....128.1623B} in the figure. The agreement is excellent. In the second panel of Fig.~\ref{fig:feh} we show the comparisons between our metallicities and those from \citet{2011AJ....141...61C} and \citet{2009A&A...508.1285G}, who derived metallicities of old globular clusters in M31 from Lick indices relations. Our results are in good agreement with the two works, with the means and dispersions of the differences of about 0.1 and 0.4, respectively. In the third panel of Fig.~\ref{fig:feh} we show the comparisons of our metallicities and those from \citet{2016AJ....152...45C} and \citet{2013A&A...549A..60C}. Metallicities from both \citet{2016AJ....152...45C} and \citet{2013A&A...549A..60C} were derived by the full spectrum fitting method based on the integrated spectra of clusters. As indicated by \citet{2016AJ....152...45C}, the full spectrum fitting method would significantly underestimate the metallicities of young clusters, which is clearly visible in the plots. For the old clusters metallicities derived from the combined fits of spectroscopic and photometric data in the current work are in agreement with those from spectral fitting in the literature  overall, except for several clusters that have metallicities from the literature around the lower metallicity limit of the SSP models.
On the left side of the bottom panels we compare our result with those from \citet{2002AJ....123.2490P}, who also obtained spectroscopic metallicities from the Lick indices. Except for those clusters that have been identified as young clusters in our work, our results  correlate well  with those from \citet{2002AJ....123.2490P}. The mean of the differences are relatively large (about 0.3\,dex). This is partly caused by the improper background subtraction for clusters near the centre of M31 in \citet{2002AJ....123.2490P}. Finally, we compare our combined fit metallicities with those estimated based on high-resolution spectroscopy \citep{2014ApJ...797..116C} on the right side of the bottom panels. The comparison shows very good agreement, with an rms of only 0.13\,dex.

Figure~\ref{fig:ages} plots the comparisons of cluster ages derived in the current work with those from the literature. We first compare our ages with those from \citet{2009AJ....137...94C} and \citet{2011AJ....141...61C}. The agreement is very good for both the young and old clusters. Our ages of four clusters (B316-G040, B349, B089D, and B240D-D066), which are found to be young by \citet{2009AJ....137...94C}, are overestimated. They are classified as old clusters by our random forest classifiers. We have checked their positions in the colour-colour and line index-index diagrams (Fig.~\ref{fig:outliers}). Two of them (B240D-D066 and B349) have blue colours like the young clusters. Their line indices locate them at the region where there are both young and old clusters. They should thus be misclassified by our random forest classifiers.  However, for the other two clusters (B316-G040 and B089D), either their colours or their line indices are similar to those of  old clusters.  On the right side of the top panels of Fig.~\ref{fig:ages}, we compare our ages with those from \citet{2004AJ....128.1623B}, who derived ages of M31 clusters by fitting the Lick indices with those from the BC03 models. The comparison shows very good agreement. The second panel of Fig.~\ref{fig:ages} show the comparisons of our age estimates with those from \citet{2009A&A...507.1375P} and \citet{2012ApJS..199...37K}, who have derived ages of M31 young clusters. Overall, our results are  consistent  with those from Perina et al. and Kang et al. There are five outliers, B347-G154, B083-G146, B316-G040, B089D, and SK007A.  Two of them (B316-G040 and B089D) are   discussed in the Caldwell et al. comparison. In Fig.~\ref{fig:outliers} we also indicate the positions of the other three clusters. SK007A is similar to the cases of B240D-D066 and B349. It could be misclassified as an old cluster by our classifiers. The other two clusters, B347-G154 and B083-G146, should be old clusters according to their red colours. In the third panel of the figure we compare our results with those from \citet{2016AJ....152...45C} and \citet{2013A&A...549A..60C}, who use the full spectral fitting method. The ages of the young clusters from \citet{2016AJ....152...45C} are also estimated from the SED fittings. They are in good agreement with our results. The ages of the old clusters from both \citet{2016AJ....152...45C} and \citet{2013A&A...549A..60C} are derived from the full spectral fitting. They span a broad range from intermediate ages to old ages. While our derived ages of the old clusters from the combined fits of both the spectroscopic and photometric data are all  very old. The ages of the young clusters from \citet{2013A&A...549A..60C} by the full spectral fitting method are systematically older than our results. On the left side of the bottom panels we compare our results with those from \citet{2010ApJ...725..200F}, who estimated the ages of M31 star clusters by the SED fitting method based on the multi-band photometric data. For young clusters our results are consistent with theirs. However, the estimates of old clusters from Fan et al. are systematically smaller than ours. The discrepancies are also significant. As mentioned before, the SED fitting ages of old clusters may have very large uncertainties as it is hard to distinguish the SED shapes of the clusters of ages larger than 1\,Gyr. Finally, on the right side of the bottom panels, we show the comparison of our results with those from \citet{2009ApJ...704..385C,2014ApJ...797..116C}, who estimated ages of clusters based on high-resolution spectroscopy. Our results are in good agreement with their estimates.

\section{Summary}

In this paper, we select confirmed M31 star clusters and candidates from the literature and match them with the LAMOST DR6. This yields a sample of 346 M31 clusters with LAMOST observations: 334  are confirmed clusters from the previous works and  12 are new clusters with radial velocity $V_{r}<-150\,\rm km\,s^{-1}$.

Based on the LAMOST spectra and the multi-band photometry of these clusters, we have developed a new algorithm to determine the parameters, including metallicities and ages of the sample clusters. We first divide our catalogue into two groups, young clusters and old clusters, by random forest classifiers; this method  can break the degeneracy between the young metal-rich and old metal-poor clusters precisely with few outliers. Based on the BATC and SDSS photometry, we perform SED analysis to the young clusters to obtain their ages. Their metallicities are estimated by fitting their spectral principle components with those from the young and metal-rich SSP models. As a result we have 53 young clusters in our catalogue. For the remaining 293 old clusters, their ages and metallicities are derived by a combined fitting of their spectroscopic and photometric data with random forest models constructed from the SSP models. The spectral principal components adopted in our method can reduce the dimensions of spectral data significantly, and the forwarding random forest models are able to fit highly non-linear relations between the observations (e.g.  the spectral principle components and the multi-band photometry) and the intrinsic parameters of clusters. Our estimated ages and metallicities are in good agreement with previous determinations in general.

\section*{Acknowledgement}

We want to thank Dr. Yinbi Li from National Astronomical Observatories of CAS for her technical support of LAMOST spectra and helpful discusses. This work is partially supported by National Natural Science Foundation of China (NSFC, Nos. 11803029 and 11873053), Yunnan University grant No. C176220100007, and National Key R\&D Program of China No. 2019YFA0405501 and 2019YFA0405503.

This work has made use of data products from the Guoshoujing Telescope (the Large Sky Area Multi-Object Fibre Spectroscopic Telescope, LAMOST). LAMOST is a National Major Scientific
Project built by the Chinese Academy of Sciences. Funding for the project has been provided by the National Development and Reform Commission. LAMOST is operated and managed by the
National Astronomical Observatories, Chinese Academy of Sciences.

\end{document}